%% file: remMap_042309.tex
\documentclass[12pt]{article}
\usepackage{graphicx,color}
\usepackage{chicago}
\usepackage{subfigure}
\usepackage{amsmath, amsfonts, amssymb, amstext}

\newtheorem{thm}{Theorem}

\definecolor{DarkBlue}{rgb}{0.1,0.1,0.5}
\definecolor{Red}{rgb}{0.9,0.0,0.1}

\usepackage{setspace}
\title{Regularized Multivariate Regression for Identifying Master
  Predictors with Application to Integrative Genomics Study of
  Breast Cancer}
\author{Jie Peng $^1$, Ji Zhu $^2$, Anna Bergamaschi $^3$, Wonshik Han$^4$, \\\vspace{15pt}
Dong-Young Noh$^4$, Jonathan R. Pollack $^5$, Pei
Wang$^{6,}$\footnote{Correspondence author: pwang@fhcrc.org}\\
$^1$Department of Statistics,  University of California, Davis, CA,
USA;\\ $^2$Department of Statistics, University of Michigan, Ann
Arbor, MI, USA;\\ $^3$Department of Genetics, Institute for Cancer
Research, Rikshospitalet-\\Radiumhospitalet Medical Center, Oslo,
Norway;\\ $^4$Cancer Research Institute and Department of Surgery,\\
Seoul National University College of Medicine, Seoul, South Korea;\\
$^5$Department of Pathology, Stanford University, CA, USA;\\
$^6$Division of Public Health Science, Fred Hutchinson Cancer
Research \\Center, Seattle, WA, USA.}

\begin{document}
\maketitle

\newpage
\begin{abstract}
In this paper, we propose a new method \texttt{remMap} ---
REgularized Multivariate regression for identifying MAster
Predictors --- for fitting multivariate response regression models
under the high-dimension-low-sample-size setting. \texttt{remMap} is
motivated by investigating the regulatory relationships among
different biological molecules based on multiple types of high
dimensional genomic data. Particularly, we are interested in
studying the influence of DNA copy number alterations on RNA
transcript levels. For this purpose, we model the dependence of the
RNA expression levels on DNA copy numbers through multivariate
linear regressions and utilize proper regularization to deal with
the high dimensionality as well as to incorporate desired network
structures. Criteria for selecting the tuning parameters are also
discussed. The performance of the proposed method is illustrated
through extensive simulation studies. Finally, \texttt{remMap} is
applied to a breast cancer study, in which genome wide RNA
transcript levels and DNA copy numbers were measured for 172 tumor
samples. We identify a trans-hub region in cytoband 17q12-q21, whose
amplification influences the RNA expression levels of more than $30$
unlinked genes. These findings may lead to a better understanding of
breast cancer pathology.

\end{abstract}



\input{Introduction_042309.tex}

\input{Method_042309.tex}

\input{Simulation_042309.tex}

\input{RealApplication_042309.tex}

\input{Discussion_042309.tex}

\bibliographystyle{chicago}

\input{remMap.bbl}
\input{Figure_032409.tex} 
\clearpage

\setcounter{page}{1}

\setcounter{table}{0}

\setcounter{figure}{0}

\setcounter{equation}{0}

\input{AppendixAB_042309.tex}
\renewcommand{\thetable}{S-\arabic{table}}
\input{AppendixC_042309.tex}
\input{remMap_Appendix.bbl}
\clearpage
\renewcommand{\thefigure}{S-\arabic{figure}}
\input{Figure_Appendix_042309.tex}

\end{document}

%% file: Introduction_042309.tex
\section{Introduction}
In a few recent breast cancer cohort studies, microarray expression
experiments and array CGH (comparative genomic hybridization)
experiments have been conducted for more than $170$ primary breast
tumor specimens collected at multiple cancer
centers~\shortcite{Sorlie:2001,Sorlie:2003,Zhao:2004,Kapp:2006,Bergamaschi:2006,Langerod:2007,Bergamaschi:2008}.
The resulting RNA transcript levels (from microarray expression
experiments) and DNA copy numbers (from CGH experiments) of
about $20K$ genes/clones across all the tumor samples were then used
to identify useful molecular markers for potential clinical usage.
While useful information has been revealed by analyzing expression
arrays alone or CGH arrays alone, careful {\it integrative analysis}
of DNA copy numbers and expression data are necessary as these two
types of data provide complimentary information in gene
characterization. Specifically, RNA data give information on genes
that are over/under-expressed, but do not distinguish primary
changes driving cancer from secondary changes resulting from cancer,
such as proliferation rates and differentiation state. On the other
hand, DNA data give information on gains and losses that are drivers
of cancer. Therefore, integrating DNA and RNA data helps to discern
more subtle (yet biologically important) genetic  regulatory
relationships in cancer cells~\shortcite{Pollack:2002}.

It is widely agreed that variations in gene copy numbers play an
important role in cancer development through altering the expression
levels of cancer-related genes~\shortcite{Albertson:2003}. This is
clear for {\it cis-regulations}, in which a gene's DNA copy number
alteration influences its own RNA transcript
level~\shortcite{Hyman:2002,Pollack:2002}. However, DNA copy number
alterations can also alter in trans the RNA transcript levels of
genes from unlinked regions, for example by directly altering the
copy number and expression of transcriptional regulators, or by
indirectly altering the expression or activity of transcriptional
regulators, or through genome rearrangements affecting
cis-regulatory elements. The functional consequences of such {\it
trans-regulations} are much harder to establish, as such inquiries
involve assessment of a large number of potential regulatory
relationships. Therefore, to refine our understanding of how these
genome events exert their effects, we need new analytical tools that
can reveal the subtle and complicated interactions among DNA copy
numbers and RNA transcript levels. Knowledge resulting from such
analysis will help shed light on cancer mechanisms.

The most straightforward way to model the dependence of RNA levels
on DNA copy numbers is through a multivariate response linear
regression model with the RNA levels being responses and the DNA
copy numbers being predictors. While the multivariate linear
regression is well studied in statistical literature, the current
problem bears new challenges due to (i) high-dimensionality in terms
of both predictors and responses; (ii) the interest in identifying
\textit{master regulators} in genetic regulatory networks; and (iii)
the complicated relationships among response variables. Thus, the
naive approach of regressing each response onto the predictors
separately is unlikely to produce satisfactory results, as such
methods often lead to high variability and over-fitting. This has
been observed by many authors, for example,
~\shortciteN{Breiman:1997} show that taking into account of the
relation among response variables helps to improve the overall
prediction accuracy. More recently, ~\shortciteN{Kim:2008} propose a
new statistical framework to explicitly incorporate the
 relationships among responses by assuming the linked responses
depend on the predictors in a similar way. The authors show that
this approach helps to select relevant predictors when the above
assumption holds.

When the number of predictors is moderate or large, model selection
is often needed for prediction accuracy and/or model interpretation.
Standard model selection tools in multiple regression such as AIC
and forward stepwise selection have been extended to multivariate
linear regression models
~\shortcite{Bedrick:1994,Fujikoshi:1997,Lutz:2006}. More recently,
sparse regularization schemes have been utilized for model selection
under the high dimensional multivariate regression setting. For
example, ~\shortciteN{Turlach:2005} propose to constrain the
coefficient matrix of a multivariate regression model to lie within
a suitable polyhedral region.~\shortciteN{Lutz:2006} propose an
$L_2$ multivariate boosting procedure. \shortciteN{Obozinskiy:2008}
propose to use a $\ell_1/\ell_2$ regularization to identify the
union support set in the multivariate regression. Moreover, Brown et
al. (1998, 1999, 2002) introduce a Bayesian framework to model the
relation among the response variables when performing variable
selection for multivariate regression. Another way to reduce the
dimensionality is through factor analysis. Related work
includes~\shortciteN{Izenman:1975},~\shortciteN{Frank:1993},~\shortciteN{Reinsel:1998},~\shortciteN{YuanMulti:2007}
and many others.

For the problem we are interested in here, the dimensions of both
predictors and responses are  large (compared to the sample size).
Thus in addition to assuming that only a subset of predictors enter
the model, it is also reasonable to assume that a predictor may
affect only some but not all responses. Moreover, in many real
applications, there often exist a subset of predictors which are
more important than other predictors in terms of model building
and/or scientific interest. For example, it is widely believed that
genetic regulatory relationships are intrinsically
sparse~\shortcite{Jeong:2001,Gardner:2003}. At the same time,  there
exist \textit{master regulators} --- network components that affect
many other components, which play important roles in shaping the
network functionality. Most methods mentioned above do not take into
account the dimensionality of the responses, and thus a
predictor/factor influences either all or none responses, e.g.,
\shortciteN{Turlach:2005}, \shortciteN{YuanMulti:2007}, the $L_2$
row boosting by \shortciteN{Lutz:2006}, and the $\ell_1/\ell_2$
regularization by \shortciteN{Obozinskiy:2008}. On the other hand,
other methods only impose a sparse model, but do not aim at
selecting a subset of predictors, e.g., the $L_2$ boosting by
\shortciteN{Lutz:2006}. In this paper, we propose a novel method
\texttt{remMap}
--- REgularized Multivariate regression for identifying MAster
Predictors, which takes into account both aspects. \texttt{remMap}
uses an $\ell_1$ norm penalty to control the overall sparsity of the
coefficient matrix of the multivariate linear regression model. In
addition, \texttt{remMap} imposes a ``group" sparse penalty, which
in essence is the same as the ``group lasso" penalty proposed by
\shortciteN{Bakin:1999}, \shortciteN{Antoniadis:2001},
\shortciteN{YuanLin:2006}, \shortciteN{Zhao:2006} and
\shortciteN{Obozinskiy:2008} (see more discussions in Section
\ref{sec:method}).  This penalty puts a constraint on the $\ell_2$
norm of regression coefficients for each predictor, which controls
the total number of predictors entering the model, and consequently
facilitates the detection of \textit{master predictors}. The
performance of the proposed method is illustrated through extensive
simulation studies.

We apply the \texttt{remMap} method on the breast cancer data set
mentioned earlier and identify a significant trans-hub region in
cytoband 17q12-q21, whose amplification influences the RNA levels of
more than 30 unlinked genes. These findings may shed some light on
breast cancer pathology. We also want to point out that analyzing
CGH arrays and expression arrays together reveals only a small
portion of the regulatory relationships among genes.  However, it
should identify many of the important relationships, i.e., those
reflecting primary genetic alterations that drive cancer development
and progression. While there are other mechanisms to alter the
expression of master regulators, for example by DNA mutation or
methylation, in most cases one should also find corresponding DNA
copy number changes in at least a subset of cancer cases.
Nevertheless, because we only identify the subset explainable by
copy number alterations, the words ``regulatory network" (``master
regulator") used in this paper will specifically refer to the
subnetwork (hubs of the subnetwork) whose functions change with DNA
copy number alterations, and thus can be detected by analyzing CGH
arrays together with expression arrays.

The rest of the paper is organized as follows. In Section 2, we
describe the \texttt{remMap} model, its implementation and criteria
for tuning. In Section 3, the performance of \texttt{remMap} is
examined through extensive simulation studies. In Section 4, we
apply the \texttt{remMap} method on the breast cancer data set. We
conclude the paper with discussions in Section 5. Technical details
are provided in the supplementary material.

%% file: Method_042309.tex
\section{Method}\label{sec:method}
\subsection{Model}
Consider multivariate regression with $Q$ response variables
$y_1,\cdots, y_Q$ and $P$ prediction variables $x_1,\cdots, x_P$:
\begin{eqnarray}
\label{eqn:mr} y_q=\sum_{p=1}^P x_p \beta_{pq}+\epsilon_q, \quad
q=1,\cdots,Q,
\end{eqnarray}
where the error terms $\epsilon_1,\cdots,\epsilon_Q$ have a joint
distribution with mean $0$ and covariance
$\mathbf{\Sigma}_{\epsilon}$. The primary goal of this paper is to
identify non-zero entries in  the $P \times Q$ coefficient matrix
$\mathbf{B}=(\beta_{pq})$ based on $N$ i.i.d samples from the above
model. Under normality assumptions, $\beta_{pq}$ can be interpreted
as proportional to the conditional correlation
$\text{Cor}(y_q,x_p|x_{-(p)})$, where
$x_{-(p)}:=\{x_{p^{\prime}}: 1 \leq p^{\prime} \neq p \leq P\}$. In
the following, we use $Y_q=(y^1_q,\cdots,y^N_q)^T$ and
$X_p=(x^1_p,\cdots,x^N_p)^T$ to denote the sample of the $q^{th}$
response variable and that of the $p^{th}$ prediction variable,
respectively. We also use $\mathbf{Y}=(Y_1:\cdots:Y_Q)$ to denote
the $N \times Q$ response matrix, and use $\mathbf{X}=(X_1:\cdots:
X_P)$ to denote the $N \times P$ prediction matrix.

In this paper, we shall focus on the cases where both $Q$ and $P$
are larger than the sample size $N$. For example, in the breast
cancer study discussed in Section~\ref{sec:application}, the sample
size is $172$, while the number of genes and the number of
chromosomal regions are on the order of a couple of hundreds (after
pre-screening). When $P>N$, the ordinary least square solution is
not unique, and regularization becomes indispensable. The choice of
suitable regularization depends heavily on the type of data
structure we envision. In recent years, $\ell_1$-norm based sparsity
constraints such as \textit{lasso} ~\shortcite{Lasso:1996} have been
widely used under such high-dimension-low-sample-size settings. This
kind of regularization is particularly suitable for the study of
genetic pathways, since genetic regulatory relationships are widely
believed to be intrinsically
sparse~\shortcite{Jeong:2001,Gardner:2003}. In this paper, we impose
an $\ell_1$ norm penalty on the coefficient matrix $\mathbf{B}$ to
control the overall sparsity of the multivariate regression model.
In addition, we put constraints on the total number of predictors
entering the model. This is achieved by treating  the coefficients
corresponding to the same predictor (one row of $\mathbf{B}$) as a
group, and then penalizing their $\ell_2$ norm. A predictor will not
be selected into the model if the corresponding $\ell_2$ norm is
shrunken to $0$. Thus this penalty facilitates the identification of
\textit{master predictors} --- predictors which affect (relatively)
many response variables. This idea is motivated by the fact that
master regulators exist and are of great interest in the study of
many real life networks including genetic regulatory networks.
Specifically, for model (\ref{eqn:mr}), we propose the following
criterion
\begin{eqnarray}
\label{eqn:loss}
L(\mathbf{B};\lambda_1,\lambda_2)=\frac{1}{2}||\mathbf{Y}-\sum_{p=1}^P
X_p B_p ||^2_F+\lambda_1\sum_{p=1}^P ||C_p\cdot B_p||_1+\lambda_2 \sum_{p=1}^P ||C_p\cdot
B_p||_2,
\end{eqnarray}
where $C_p$ is the $p$th row of
$\mathbf{C}=(c_{pq})=(C_1^T:\cdots:C_P^T)^T$, which is a
pre-specified $P \times Q$ 0-1 matrix indicating the coefficients on
which penalization is imposed; $B_p$ is the $p^{th}$ row of
$\mathbf{B}$; $||\cdot||_F$ denotes the Frobenius norm of matrices;
$||\cdot||_1$ and $||\cdot||_2$ are the $\ell_1$ and $\ell_2$ norms
for vectors, respectively; and ``$\cdot$" stands for Hadamard
product (that is, entry-wise multiplication). The indicator matrix
$\mathbf{C}$ is pre-specified based on prior knowledge: if we know
in advance that predictor $x_p$ affects response $y_q$, then the
corresponding regression coefficient $\beta_{pq}$ will not be
penalized and we set $c_{pq}=0$ (see Section \ref{sec:application}
for an example). When there is no such prior information,
$\mathbf{C}$ can be simply set to a constant matrix $c_{pq} \equiv
1$. Finally, an estimate of the coefficient matrix $\mathbf{B}$ is
$\widehat{\mathbf{B}}(\lambda_1,\lambda_2):=\arg \min_{\mathbf{B}}
L(\mathbf{B};\lambda_1,\lambda_2)$.

In the above criterion function, the $\ell_1$ penalty induces the
overall sparsity of the coefficient matrix $\mathbf{B}$. The
$\ell_2$ penalty on the row vectors $C_p\cdot B_p$ induces row
sparsity of the product matrix $\mathbf{C}\cdot \mathbf{B}$. As a
result, some rows are shrunken to be entirely zero (Theorem
\ref{thm:fitting}). Consequently, predictors which affect relatively
more response variables are more likely to be selected into the
model. We refer to the combined penalty in equation~(\ref{eqn:loss})
as the \texttt{MAP} (MAster Predictor) penalty. We also refer to the
proposed estimator $\widehat{\mathbf{B}}(\lambda_1,\lambda_2)$ as
the \texttt{remMap} (REgularized Multivariate regression for
identifying MAster Predictors) estimator. Note that, the $\ell_2$
penalty is a special case (with $\alpha=2$) of the more general
penalty form: $ \sum_{p=1}^P ||C_p\cdot B_p||_\alpha$, where
$||v||_\alpha:=(\sum_{q=1}^Q |v_q|^{\alpha})^{\frac{1}{\alpha}}$ for
a vector $v \in \mathcal{R}^Q$ and $\alpha>1$. In
~\shortciteN{Turlach:2005}, a penalty with $\alpha=\infty$ is used
to select a common subset of prediction variables when modeling
multivariate responses. In Yuan et al. (2007), a constraint with
$\alpha=2$ is applied to the loading matrix in a multivariate linear
factor regression model for dimension reduction. In
~\shortciteN{Obozinskiy:2008}, the same constraint is applied to
identify the union support set in the multivariate regression. In
the case of multiple regression, a similar penalty corresponding to
$\alpha=2$ is proposed by~\shortciteN{Bakin:1999} and
by~\shortciteN{YuanLin:2006} for the selection of grouped variables,
which corresponds to the blockwise additive penalty in
\shortciteN{Antoniadis:2001} for wavelet shrinkage.
\shortciteN{Zhao:2006} propose the penalty with a general
$\alpha>1$. However, none of these methods takes into account the
high dimensionality of response variables and thus
predictors/factors are simultaneously selected for all responses. On
the other hand, by combining the $\ell_2$ penalty and the $\ell_1$
penalty together in the \texttt{MAP} penalty, the \texttt{remMap}
model not only selects a subset of predictors, but also limits the
influence of the selected predictors to only some (but not
necessarily all) response variables. Thus, it is more suitable for
the cases when both the number of predictors and the number of
responses are large. Lastly, we also want to point out a difference
  between the \texttt{MAP} penalty and the \texttt{ElasticNet}
penalty proposed by ~\shortciteN{Zou:2005}, which combines the
$\ell_1$ norm penalty with the squared $\ell_2$ norm penalty. The
\texttt{ElasticNet} penalty aims to encourage a group selection
effect for highly correlated predictors under the multiple
regression setting. However, the squared $\ell_2$ norm itself does
not induce sparsity and thus is intrinsically different from the
$\ell_2$ norm penalty discussed above.

In Section \ref{sec:simulation}, we use extensive simulation studies
to illustrate the effects of the \texttt{MAP} penalty. We compare
the \texttt{remMap} method with two alternatives: (i) the
\texttt{joint} method which only utilizes the $\ell_1$ penalty, that
is $\lambda_2=0$ in (\ref{eqn:loss}); (ii) the \texttt{sep} method
which performs $Q$ separate lasso regressions. We find that, if
there exist large hubs (master predictors), \texttt{remMap} performs
much better than \texttt{joint} in terms of identifying the true
model; otherwise, the two methods perform similarly. This suggests
that the ``simultaneous" variable selection enhanced by the $\ell_2$
penalty pays off when there exist a small subset of ``important"
predictors, and it costs little when such predictors are absent. In
addition,  both \texttt{remMap} and \texttt{joint} methods impose
sparsity of the coefficient matrix as a whole. This helps to borrow
information across different regressions corresponding to different
response variables, and thus incorporates the relationships among
response variables into the model. It also amounts to a greater
degree of regularization, which is usually desirable for the
high-dimension-low-sample-size setting. On the other hand, the
\texttt{sep} method controls sparsity for each individual regression
separately and thus is subject to high variability and over-fitting.
As can be seen by the simulation studies (Section
\ref{sec:simulation}), this type of ``joint" modeling greatly
improves the model efficiency. This is also noted by other authors
including ~\shortciteN{Turlach:2005}, ~\shortciteN{Lutz:2006} and
~\shortciteN{Obozinskiy:2008}.



\subsection{Model Fitting}
In this section, we propose an iterative algorithm for solving the
\texttt{remMap} estimator $\widehat{\mathbf{B}}(\lambda_1,
\lambda_2)$. This is a convex optimization problem when the two
tuning parameters are not both zero, and thus there exists a unique
solution. We first describe how to update one row of
$\mathbf{\mathbf{B}}$, when all other rows are fixed.

{\thm \label {thm:fitting} Given $\{B_p\}_{p \not=p_0}$ in
(\ref{eqn:loss}), the solution for $
\min_{B_{p_0}}L(\mathbf{B};\lambda_1,\lambda_2)$ is given by
$\widehat{B}_{p_0}=(\widehat{\beta}_{p_0,1},\cdots,\widehat{\beta}_{p_0,Q})$
which satisfies: for $1 \leq q \leq Q$

\begin{itemize}
\item [(i)] If $c_{p_0,q}=0$,
$\widehat{\beta}_{p_0,q}=X_{p_0}^T\widetilde{Y}_{q}/\|X_{p_0}\|_2^2$
(OLS), where $\widetilde{Y}_{q}=Y_q-\sum_{p\neq p_0} X_p\beta_{pq}$;
\item [(ii)] If $c_{p_0,q}=1$,
\begin{eqnarray}
\label{eqn:group} \widehat{\beta}_{p_0,q} =\left\{
\begin{array}{ll}
  0, &{\textrm if}~ \| \widehat{B}_{p_0}^{{\textrm lasso}} \|_{2,
    C}=0;\\
  \left(1-\frac{\lambda_2}{\|\widehat{\mathbf{B}}_{p_0}^{{\textrm
    lasso}}\|_{2,
      C}\cdot
    \|X_{p_0}\|_2^2}\right)_{+}\widehat{\beta}_{p_0,q}^{\rm{lasso}},
  & \rm{otherwise,} \end{array} \right.
\end{eqnarray}
\end{itemize}
where $$||\widehat{B}_{p_0}^{\rm{lasso}}||_{2,
C}:=\left\{\sum_{q=1}^Q
c_{p_0,q}(\widehat{\beta}_{p_0,q}^{\rm{lasso}})^2 \right\}^{1/2},$$
and
\begin{eqnarray}
\label{eqn:lasso} \widehat{\beta}_{p_0,q}^{\rm{lasso}}=\left\{
\begin{array}{ll}
X_{p_0}^T\widetilde{Y}_{q}/\|X_{p_0}\|_2^2, &{\textrm if}~ c_{p_0,q}=0;\\
\left(|X_{p_0}^T\widetilde{Y}_{q}|-\lambda_1\right)_{+}
\frac{{\textrm sign}\left(X_{p_0}^T\widetilde{Y}_{q}\right)}{\|X_{p_0}\|_2^2},
&{\textrm if}~ c_{p_0,q}=1.\end{array}\right.
\end{eqnarray}
}
The proof of Theorem 1 is given in the supplementary material (Appendix A).

Theorem \ref{thm:fitting} says that, when estimating the $p_0^{th}$
row of the coefficient matrix $\mathbf{B}$ with all other rows
fixed: if there is a pre-specified relationship between the
$p_0^{th}$ predictor and the $q^{th}$ response (i.e.,
$c_{p_0,q}=0$), the corresponding coefficient $\beta_{p_0,q}$ is
estimated by the (univariate) ordinary least square solution (OLS)
using current responses $\widetilde{Y}_q$; otherwise, we first
obtain the lasso solution $\widehat{\beta}_{p_0,q}^{\rm{lasso}}$ by
the (univariate) soft shrinkage of the OLS solution (equation
(\ref{eqn:lasso})), and then conduct a group shrinkage of the lasso
solution (equation (\ref{eqn:group})). From Theorem
\ref{thm:fitting}, it is easy to see that, when the design matrix
$\mathbf{X}$ is orthonormal: $\mathbf{X}^T\mathbf{X}=I_p$ and
$\lambda_1=0$, the \texttt{remMap} method amounts to selecting
variables according to the $\ell_2$ norm of their corresponding OLS
estimates.

Theorem \ref{thm:fitting} naturally leads to an algorithm which
updates the rows of $\mathbf{B}$ iteratively until convergence. In
particular, we adopt the \texttt{active-shooting} idea proposed
by~\shortciteN{space:2008} and ~\shortciteN{Friedman:2008}, which is
a modification of the \texttt{shooting}
algorithm proposed by ~\shortciteN{Fu:1998} and also
~\shortciteN{Friedman:2007} among others. The algorithm proceeds
as follows:
\begin{itemize}
\item[1.] Initial step: for $p=1, ..., P$; $q=1, ..., Q$,
\begin{eqnarray}
\label{eqn:remMap:initial} \widehat{\beta}_{p,q}^{0}=\left\{
\begin{array}{lll}
X_{p}^T Y_{q}/\|X_{p}\|_2^2, &\rm{if} & c_{p,q}=0;\\
\left(|X_{p}^T Y_{q}|-\lambda_1\right)_{+}
\frac{{\textrm sign}\left(X_{p}^T Y_{q}\right)}{\|X_{p}\|_2^2},
&\rm{if} & c_{p,q}=1.\end{array}\right.
\end{eqnarray}
\item[2.]Define the current {\it active-row set} $\Lambda=\{p:
\textrm{ current }||\widehat{B}_{p}||_{2, C}\neq 0 \}$.
\begin{itemize}
\item[(2.1)] For each $p\in \Lambda$, update $\widehat{B}_{p}$
with all other rows of $\mathbf{B}$ fixed at their current values according to
Theorem \ref{thm:fitting}. \item[(2.2)] Repeat (2.1) until
convergence is achieved on the current active-row set.
\end{itemize}
\item[3.]For $p=1 \textrm{ to } P$, update $\widehat{B}_{p}$ with
all other rows of $\mathbf{B}$ fixed at their current values according to
Theorem \ref{thm:fitting}. If no $\widehat{B}_{p}$ changes during
this process, return the current $\widehat{\mathbf{B}}$ as the final
estimate. Otherwise, go back to step 2.
\end{itemize}
It is clear that the computational cost of the above algorithm is in
the order of $O(NPQ)$.

\subsection{Tuning}\label{sec:tuning}

In this section, we discuss the selection of the tuning parameters
$(\lambda_1,\lambda_2)$ by v-fold cross validation.
To perform the v-fold cross validation, we first partition the whole
data set into $V$ non-overlapping subsets, each consisting of
approximately $1/V$ fraction of total samples. Denote the $i^{th}$
subset as $D^{(i)}=(\mathbf{Y}^{(i)}, \mathbf{X}^{(i)})$, and its
complement as $D^{-(i)}=(\mathbf{Y}^{-(i)}, \mathbf{X}^{-(i)})$. For
a given $(\lambda_1, \lambda_2)$, we obtain the \texttt{remMap}
estimate:
$\widehat{\mathbf{B}}^{(i)}(\lambda_1,\lambda_2)=(\widehat{\beta}_{pq}^{(i)})$
based on the $i^{th}$ training set $D^{-(i)}$. We then obtain the
\textit{ordinary least square estimates}
$\widehat{\mathbf{B}}_{\rm{ols}}^{(i)}(\lambda_1,\lambda_2) =
(\widehat{\beta}_{\textrm{ols},pq}^{(i)})$ as follows: for $1 \leq q
\leq Q$, define $S_q=\{p: 1 \leq p \leq P,
\widehat{\beta}_{pq}^{(i)} \neq 0 \}$. Then set
$\widehat{\beta}_{\textrm{ols},pq}^{(i)}=0$ if $p \notin S_q$;
otherwise, define $\{\widehat{\beta}_{\textrm{ols},pq}^{(i)}: p \in
S_q \}$ as the ordinary least square estimates by regressing
$Y_q^{-(i)}$ onto $\{X_p^{-(i)}: p \in S_q\}$. Finally, prediction
error is calculated on the test set $D^{(i)}$:
\begin{eqnarray}
\label{eqn:cv_i}
\texttt{remMap.cv}_i(\lambda_1,\lambda_2):=||\mathbf{Y}^{(i)}-\mathbf{X}^{(i)}\widehat{\mathbf{B}}_{\rm{ols}}^{(i)}(\lambda_1,\lambda_2)||^2_2.
\end{eqnarray}
The v-fold cross validation score is then defined as
\begin{eqnarray}
\label{eqn:cv} \texttt{remMap.cv}(\lambda_1,\lambda_2)=\sum_{i=1}^V
\texttt{remMap.cv}_i(\lambda_1,\lambda_2).
\end{eqnarray}

The reason for using OLS estimates in calculating the prediction error
is because the true model is assumed to be sparse. As noted by
\shortciteN{lars}, when there are many noise variables, using
shrunken estimates in the cross validation criterion often results
in over fitting. Similar results are observed in our simulation
studies: if in (\ref{eqn:cv_i}) and (\ref{eqn:cv}), the shrunken
estimates are used, the selected models are all very big which
result in large numbers of false positive findings.
In addition, we
also try AIC and GCV  for tuning and both criteria result in over
fitting as well. These results are not reported  in the next section
due to space limitation.

In order to further control the false positive findings, we propose
a method called \texttt{cv.vote}. The idea is to treat the training
data from each cross-validation fold as a ``bootstrap" sample. Then
variables being consistently selected by many cross validation folds
should be more likely to appear in the true model than the variables
being selected only by few cross validation folds. Specifically, for
$1\leq p \leq P$ and $1\leq q \leq Q$, define
\begin{eqnarray}\label{eqn:cv.vote}
 s_{pq}(\lambda_1, \lambda_2)=
 \left\{
 \begin{array}{ll}
1, & \textrm{ if } \sum_{i=1}^V
I(\widehat{\beta}_{pq}^{(i)}(\lambda_1, \lambda_2)\neq 0) > V_a;\\
0, & \textrm{ otherwise}.
\end{array}\right.
\end{eqnarray}
where $V_a$ is a pre-specified integer. We then select edge $(p,q)$
if $s_{pq}(\lambda_1, \lambda_2) =1 $. In the next section, we use
$V_a=5$ and thus \texttt{cv.vote} amounts to a ``majority vote"
procedure. Simulation studies in Section~\ref{sec:simulation}
suggest that, \texttt{cv.vote} can effectively decrease the number
of false positive findings while only slightly increase the number
of false negatives.

An alternative tuning method is by a BIC criterion. Compared to
v-fold cross validation, BIC is computationally cheaper. However it
requires much more assumptions. In particular, the BIC method uses
the degrees of freedom of each \texttt{remMap} model which is
difficult to estimate in general. In the supplementary material, we
derive an unbiased estimator for the degrees of freedom of the
\texttt{remMap} models when the predictor matrix $\mathbf{X}$ has
orthogonal columns (Theorem 2 of Appendix B in the supplementary
materials). In Section \ref{sec:simulation}, we show by extensive
simulation studies that, when the correlations among the predictors
are complicated, this estimator tends to select very small models.
For more details see the supplementary material, Appendix B.

%% file: Simulation_042309.tex
\section{Simulation}

\label{sec:simulation} In this section, we investigate the
performance of the \texttt{remMap} model and compare it with two
alternatives: (i) the \texttt{joint} model with $\lambda_2=0$ in
(\ref{eqn:loss}); (ii) the \texttt{sep} model which performs $Q$
separate lasso regressions. For each model, we consider three tuning
strategies, which results in nine methods in total:
\begin{enumerate}\vspace*{-2pt}
\item \texttt{remMap.cv, joint.cv, sep.cv}:
  \hspace{10pt}The tuning parameters are selected through 10-fold
  cross validation;
\vspace*{-10pt}
\item \texttt{remMap.cv.vote, joint.cv.vote, sep.cv.vote}:
  \hspace{10pt}The \texttt{cv.vote} procedure with
$V_a=5$ is applied to the models resulted from the corresponding
\texttt{$*$.cv} approaches; \vspace*{-10pt}
\item \texttt{remMap.bic, joint.bic, sep.bic}:
  \hspace{10pt} The tuning parameters are selected by a BIC
criterion. For \texttt{remMap.bic} and \texttt{joint.bic}, the
degrees of freedom are estimated according to equation
(\ref{eqn:df_rem_q}) in Appendix B of the supplementary material;
for \texttt{sep.bic}, the degrees of freedom of each regression is
estimated by the total number of selected predictors
~\shortcite{Zou:2007}. \vspace*{-10pt}
\end{enumerate}

We simulate data as follows. Given $(N,P,Q)$, we first generate the
predictors $(x_1,\cdots,x_P)^T \sim
~{\textrm Normal}_P(0,\mathbf{\Sigma}_X)$, where $\mathbf{\Sigma}_X$ is
the predictor covariance matrix (for simulations 1 and 2,
$\mathbf{\Sigma}_X(p,p^{\prime}):=\rho_x^{|p-p^{\prime}|}$). Next,
we simulate a $P \times Q$ 0-1 adjacency matrix $\mathbf{A}$, which
specifies the topology of the network between predictors and
responses, with $\mathbf{A}(p,q)=1$ meaning that $x_p$ influences
$y_q$, or equivalently $\beta_{pq} \neq 0$.  In all simulations, we
set $P=Q$ and the diagonals of $\mathbf{A}$ equal to one, which is
viewed as prior information (thus the diagonals of $\mathbf{C}$ are
set to zero). This aims to mimic \texttt{cis-regulations} of DNA
copy number alternations on its own expression levels. We then
simulate the $P \times Q$ regression coefficient matrix
$\mathbf{B}=(\beta_{pq})$ by setting $\beta_{pq}=0$, if
$\mathbf{A}(p,q)=0$; and $\beta_{pq}\sim {\rm Uniform}([-5,-1] \cup
[1,5])$, if $\mathbf{A}(p,q)=1$. After that, we generate the
residuals $(\epsilon_1,\cdots,\epsilon_Q)^T \sim
~{\textrm Normal}_Q(0,\mathbf{\Sigma}_{\epsilon})$, where
$\mathbf{\Sigma}_{\epsilon}(q,q^{\prime})=\sigma^2_{\epsilon}
\rho_{\epsilon}^{|q-q^{\prime}|}$. The residual variance
$\sigma^2_{\epsilon}$ is chosen such that the average signal to
noise ratio equals to a pre-specified level $s$. Finally, the
responses $(y_1,\cdots,y_Q)^T$ are generated according to model
(\ref{eqn:mr}). Each data set consists of  $N$ i.i.d samples of such
generated predictors and responses.  For all methods, predictors and
responses are standardized to have (sample) mean zero and standard
deviation one before model fitting. Results reported for each
simulation setting are averaged over $25$ independent data sets.

For all simulation settings, $\mathbf{C}=(c_{pq})$ is taken to be
$c_{pq}=0$, if $p=q$; and $c_{pq}=1$, otherwise. Our primary goal is
to identify the \texttt{trans-edges} --- the predictor-response pairs
$(x_p,y_q)$ with $\mathbf{A}(p,q) =1$ and $\mathbf{C}(p,q)=1$, i.e.,
the edges that are not pre-specified by the indicator matrix
$\mathbf{C}$. Thus, in the following, we report the number of false
positive detections of \texttt{trans-edges} (FP) and the number of
false negative detections of \texttt{trans-edges} (FN) for each
method. We also examine these methods in terms of predictor
selection. Specifically, a predictor is called a
\texttt{cis-predictor} if it does not have any \texttt{trans-edges};
otherwise it is called a \texttt{trans-predictor}. Moreover, we say
a false positive trans-predictor (FPP) occurs if a
\texttt{cis-predictor} is incorrectly identified as a
\texttt{trans-predictor}; we say a false negative trans-predictor
(FNP) occurs if it is the other way around.

\subsection*{Simulation I}
We first assess the performances of the nine methods under various
combinations of model parameters. Specifically, we consider:
$P=Q=400,600,800$; $s=0.25,0.5,0.75$; $\rho_x=0,0.4,0.8$; and
$\rho_{\epsilon}=0,0.4,0.8$. For all settings, the sample size $N$
is fixed at $200$. The networks (adjacency matrices $\mathbf{A}$)
are generated with $5$ master predictors (hubs), each influencing
$20 \sim 40$ responses; and all other predictors are
\texttt{cis-predictors}. We set the total number of
\texttt{tran-edges} to be $132$ for all networks. Results on
\texttt{trans-edge} detection are summarized in Figures \ref{Fig:1}
and \ref{Fig:2}.
 From these figures, it is clear that \texttt{remMap.cv}
and \texttt{remMap.cv.vote} perform the best in terms of the total
number of false detections (FP+FN), followed by \texttt{remMap.bic}.
The three \texttt{sep} methods result in too many false positives
(especially \texttt{sep.cv}). This is expected since there are in
total $Q$ tuning parameters selected separately, and the relations
among responses are not utilized at all. This leads to high
variability and over-fitting. The three \texttt{joint} methods
perform reasonably well, though they have considerably larger number
of false negative detections compared to \texttt{remMap} methods.
This is because the \texttt{joint} methods incorporate less
information about the relations among the responses caused by the
master predictors. Finally, comparing \texttt{cv.vote} to
\texttt{cv}, we can see that the \texttt{cv.vote} procedure
effectively decreases the false positive detections and only
slightly inflates the false negative counts.

As to the impact of different model parameters, signal size $s$
plays an important role for all methods: the larger the signal size,
the better these methods perform (Figure \ref{Fig:signal}).
 Dimensionality $(P,Q)$ also shows consistent impacts on
these methods: the larger the dimension, the more false negative
detections (Figure \ref{Fig:dim}).
 With increasing predictor correlation $\rho_x$, both
\texttt{remMAP.bic} and \texttt{joint.bic} tend to select smaller
models, and consequently result in less false positives and more
false negatives (Figure \ref{Fig:corr}). This is because when the
design matrix $\mathbf{X}$ is further away from orthogonality,
(\ref{eqn:df_rem_q}) tends to overestimate the degrees of freedom
and consequently smaller models are selected. The residual
correlation $\rho_{\epsilon}$ seems to have little impact on
\texttt{joint} and \texttt{sep}, and some (though rather small)
impacts on \texttt{remMap} (Figure \ref{Fig:corr_res}).
Moreover, \texttt{remMap} performs much better than \texttt{joint}
and \texttt{sep} on master predictor selection, especially in terms
of the number of false positive \texttt{trans-predictors} (results
not shown). This is because the $\ell_2$ norm penalty is more
effective than the $\ell_1$ norm penalty in excluding irrelevant
predictors.

\subsection*{Simulation II}
In this simulation, we study the performance of these methods on a
network without big hubs. The data are generated similarly as before
with $P=Q=600$, $N=200$, $s=0.25$, $\rho_x=0.4$, and
$\rho_\epsilon=0$. The network consists of $540$
\texttt{cis-predictors}, and $60$ \texttt{trans-predictors} with $1
\sim 4$ \texttt{trans-edges}. This leads to $151$
\texttt{trans-edges} in total.  As can be seen from Table
\ref{table:comp_five_pois}, \texttt{remMap} methods and
\texttt{joint} methods now perform very similarly and both are
considerably better than the \texttt{sep} methods. Indeed, under
this setting, $\lambda_2$ is selected (either by \texttt{cv} or
\texttt{bic}) to be small in the \texttt{remMap} model, making it
very close to the \texttt{joint} model.
\begin{table}[h]
\small \centering \caption{Simulation II. Network
 topology: uniform network with $151$ \texttt{trans-edges} and
 $60$ \texttt{trans-predictors}.
 $P=Q=600, N=200$; $s=0.25$; $\rho_x=0.4$; $\rho_\epsilon=0$. }
\begin{tabular}{c|ccc|cc}\hline\hline
Method & FP & FN &TF &FPP & FNP
\\\hline
\texttt{remMap.bic}&4.72(2.81) &45.88(4.5)
&50.6(4.22)&1.36(1.63)&11(1.94)\\
\texttt{remMap.cv}&18.32(11.45) &40.56(5.35)
&58.88(9.01)&6.52(5.07)&9.2(2)\\
\texttt{remMap.cv.vote}&2.8(2.92) & 50.32(5.38) & 53.12(3.94)&
0.88(1.26)& 12.08(1.89)\\\hline \texttt{joint.bic}&5.04(2.68)
&52.92(3.6) &57.96(4.32)&4.72(2.64)&9.52(1.66)\\
\texttt{joint.cv}&16.96(10.26) &46.6(5.33) &63.56(7.93) &
15.36(8.84)&7.64(2.12)\\ \texttt{joint.cv.vote}& 2.8(2.88)
&56.28(5.35) &59.08(4.04) & 2.64(2.92)&10.40(2.08)\\\hline
\texttt{sep.bic}&78.92(8.99)
&37.44(3.99) &116.36(9.15)&67.2(8.38)&5.12(1.72)\\
\texttt{sep.cv}& 240.48(29.93) &32.4(3.89)&272.88(30.18)&179.12(18.48)&2.96(1.51)\\
\texttt{sep.cv.vote}& 171.00(20.46) & 33.04(3.89)& 204.04(20.99)& 134.24(14.7)&3.6(1.50)\\
\hline\hline
\multicolumn{6}{l}{FP: \textit{false positive};\hspace{3pt}FN: \textit{false negative};\hspace{3pt}TF: \textit{total false};\hspace{3pt}FPP: \textit{false positive trans-predictor};}\\
\multicolumn{6}{l}{ FNP: \textit{false negative trans-predictor}.\hspace{3pt}\textit{Numbers in the parentheses are standard deviations}}\\
\end{tabular}
\label{table:comp_five_pois}
\end{table}

\subsection*{Simulation III}
In this simulation, we try to mimic the true predictor covariance
and network topology in the real data discussed in the next section.
We observe that, for chromosomal regions on the same chromosome, the
corresponding copy numbers are usually positively correlated, and
the magnitude of the correlation decays slowly with genetic
distance. On the other hand, if two regions are on different
chromosomes, the correlation between their copy numbers could be
either positive or negative and in general the magnitude is much
smaller than that of the regions on the same chromosome. Thus in
this simulation, we first partition the $P$ predictors into $23$
distinct blocks, with the size of the $i^{th}$ block proportional to
the number of CNAI (copy number alteration intervals) on the
$i^{th}$ chromosome of the real data (see Section
\ref{sec:application} for the definition of CNAI). Denote the
predictors within the $i^{th}$ block as $x_{i1},\cdots,x_{ig_i}$,
where $g_i$ is the size of the $i^{th}$ block. We then define the
\textit{within-block} correlation as: ${\rm
Corr}(x_{ij},x_{il})=\rho_{{\rm wb}}^{0.5|j-l|}$ for $1 \leq j,l
\leq g_i$; and define the \textit{between-block} correlation as
${\rm Corr}(x_{ij,},x_{kl}) \equiv \rho_{ik}$ for $1 \leq j \leq
g_i, \ 1 \leq l \leq g_k$ and $1 \leq i \not=k \leq 23$. Here,
$\rho_{ik}$ is determined in the following way: its sign is randomly
generated from $\{-1,1\}$; its magnitude is randomly generated from
$\{\rho_{{\rm bb}},\rho_{{\rm bb}}^2,\cdots,\rho_{{\rm bb}}^{23}\}$.
In this simulation, we set $\rho_{{\rm wb}}=0.9$, $\rho_{{\rm
bb}}=0.25$ and use $P=Q=600, N=200$, $s=0.5$, and
$\rho_{\epsilon}=0.4$. The heatmaps of the (sample) correlation
matrix of the predictors in the simulated data and that in the real
data are given by Figure S-2 in the supplementary material. The
network is generated with five large hub predictors each having $14
\sim 26 $ \texttt{trans-edges}; five small hub predictors each
having $3 \sim 4$ \texttt{trans-edges}; $20$ predictors having $1
\sim 2$ \texttt{trans-edges}; and all other predictors being
\texttt{cis-predictors}.

The results are summarized in Table
\ref{table:comp_five_pois_s0.5_151}. Among the nine methods,
\texttt{remMap.cv.vote} performs the best in terms of both edge
detectiion and master predictor prediction. \texttt{remMAP.bic} and
\texttt{joint.bic} result in very small models due to the
complicated correlation structure among the predictors. While all
three cross-validation based methods have large numbers of false
positive findings, the three \texttt{cv.vote} methods have much
reduced false positive counts and only slightly increased false
negative counts. These findings again suggest that \texttt{cv.vote}
is an effective procedure in controlling false positive rates while
not sacrificing too much in terms of power.


\begin{table}[h]
 \small \centering
\caption{Simulation III. Network topology: five large hubs
and five small hubs with $151$ \texttt{trans-edges} and $30$
\texttt{trans-predictors}. $P=Q=600,\ N=200; \ s=0.5; \ \rho_{{\rm
wb}}=0.9, \ \rho_{{\rm bb}}=0.25; \rho_\epsilon=0.4$.}
\begin{tabular}{c|ccc|cc}\hline\hline
Method & FP & FN & TF&FPP&FNP
\\\hline
\texttt{remMap.bic}&0(0) &150.24(2.11) &150.24(2.11)&0(0)&29.88(0.33) \\
\texttt{remMap.cv}&93.48(31.1) &20.4(3.35) &113.88(30.33)&15.12(6.58)&3.88(1.76)\\
\texttt{remMap.cv.vote}&48.04(17.85) &27.52(3.91) &75.56(17.67)
&9.16(4.13)&5.20(1.91)\\\hline
\texttt{joint.bic}&7.68(2.38) &104.16(3.02)&111.84(3.62)&7(2.18)&10.72(1.31)\\
\texttt{joint.cv}&107.12(13.14)&39.04(3.56)&146.16(13.61)&66.92(8.88)&1.88(1.2)\\
\texttt{joint.cv.vote}&63.80(8.98) &47.44(3.90)
&111.24(10.63)&41.68(6.29)&2.88(1.30)\\\hline
\texttt{sep.bic}&104.96(10.63)&38.96(3.48)&143.92(11.76)&64.84(6.29)&1.88(1.17)\\
\texttt{sep.cv}&105.36(11.51)&37.28(4.31)&142.64(12.26)&70.76(7.52)&1.92(1.08)\\
\texttt{sep.cv.vote}& 84.04(10.47)& 41.44(4.31)&
125.48(12.37)&57.76 (6.20)&2.4 (1.32)\\\hline\hline
\multicolumn{6}{l}{FP: \textit{false positive};\hspace{3pt}FN: \textit{false negative};\hspace{3pt}TF: \textit{total false};\hspace{3pt}FPP: \textit{false positive trans-predictor};}\\
\multicolumn{6}{l}{ FNP: \textit{false negative trans-predictor}.\hspace{3pt}\textit{Numbers in the parentheses are standard deviations}}\\
\end{tabular}
\label{table:comp_five_pois_s0.5_151}
\end{table}

%% file: RealApplication_042309.tex
\section{Real application}
\label{sec:application} In this section, we apply the proposed
\texttt{remMap} method to the breast cancer study mentioned earlier.
Our goal is to search for genome regions whose copy number
alterations have significant impacts on RNA expression levels,
especially  on those of the unlinked genes, i.e., genes not falling
into the same genome region. The findings resulting from this
analysis may help to cast light on the complicated interactions
among DNA copy numbers and RNA expression levels.

\subsection{Data preprocessing}
The $172$ tumor samples were analyzed using cDNA expression
microarray and CGH array experiments as described in
\shortciteN{Sorlie:2001},~\shortciteN{Sorlie:2003},~\shortciteN{Zhao:2004},~\shortciteN{Kapp:2006},
~\shortciteN{Bergamaschi:2006}, ~\shortciteN{Langerod:2007},
and~\shortciteN{Bergamaschi:2008}. In below, we outline the data
preprocessing steps. More details are provided in the supplementary
material (Appendix C).

Each CGH array contains measurements ($log_2$ ratios) on about $17K$
mapped human genes. A positive (negative) measurement suggests a
possible copy number gain (loss). After proper normalization,
\texttt{cghFLasso}~\shortcite{cghFLasso2008} is used to estimate the
DNA copy numbers based on array outputs. Then, we derive {\it copy
number alteration intervals} (CNAIs) --- basic CNA units (genome
regions) in which genes tend to be amplified or deleted at the same
time within one sample --- by employing the Fixed-Order Clustering
(FOC) method ~\shortcite{Wang:thesis}. In the end, for each CNAI in
each sample, we calculate the mean value of the estimated copy
numbers of the genes falling into this CNAI. This results in a $172$
(samples) by $384$ (CNAIs) numeric matrix.

Each expression array contains measurements for about $18K$ mapped
human genes. After global normalization for each array, we also
standardize each gene's measurements across 172 samples to
median$=0$ and MAD (median absolute deviation) $=1$. Then we focus
on a set of 654 breast cancer related genes, which is derived based
on 7 published breast cancer gene
lists~\shortcite{Sorlie:2003,vandeVijver:2002,Chang:2004,Paik:2004,Wang:2005,Sotiriou:2006,Saal:2007}.
This results in a $172$ (samples) by $654$ (genes) numeric matrix.

 When the copy number change of one CNAI affects the RNA
level of an unlinked gene, there are two possibilities: (i) the copy
number change directly affects the RNA level of the unlinked gene;
(ii) the copy number change first affects the RNA level of an
intermediate gene (either linked or unlinked), and then the RNA
level of this intermediate gene affects that of the unlinked gene.
Figure~\ref{figure:diagram} gives an illustration of these two
scenarios. In this study, we are more interested in finding the
relationships of the first type. Therefore, we first characterize
the interactions among RNA levels and then account for these
relationships in our model so that we can better infer direct
interactions. For this purpose, we apply the \texttt{space} (Sparse
PArtial Correlation Estimation) method to search for associated RNA
pairs through identifying non-zero partial
correlations~\shortcite{space:2008}. The estimated (concentration)
network (referred to as \textit{Exp.Net.664} hereafter) has in total
$664$ edges  --- 664 pairs of genes whose RNA levels significantly
correlate with each other after accounting for the expression levels
of other genes.

Another important factor one needs to consider when studying breast
cancer is the existence of distinct tumor subtypes. Population
stratification due to these distinct subtypes could confound our
detection of associations between CNAIs and gene expressions.
Therefore, we introduce a set of subtype indicator variables, which
later on is used as additional predictors in the \texttt{remMap}
model. Specifically, following ~\shortciteN{Sorlie:2003}, we divide
the 172 patients into 5 distinct groups based on their expression
patterns. These groups correspond to the same 5 subtypes suggested
by~\shortciteN{Sorlie:2003}
--- Luminal Subtype A, Luminal Subtype B, ERBB2-overexpressing
Subtype, Basal Subtype and Normal Breast-like Subtype.

\subsection{Interactions between CNAIs and RNA
expressions}\label{sec:real:result} We then apply the
\texttt{remMap} method to study the interactions between CNAIs and
RNA transcript levels. For each of the 654 breast cancer
genes, we regress its expression level on three sets of predictors:
(i) expression levels of other genes that are connected to the
target gene (the current response variable) in \textit{Exp.Net.664};
(ii) the five subtype indicator variables derived in the previous
section; and (iii) the copy numbers of all $384$ CNAIs. We are
interested in whether any unlinked CNAIs are selected into this
regression model (i.e., the corresponding regression coefficients
are non-zero). This suggests potential trans-regulations
(\texttt{trans-edges}) between the selected CNAIs and the target
gene expression.  The coefficients of the linked CNAI of the target
gene are not included in the \texttt{MAP} penalty (this corresponds
to $c_{pq}=0$, see Section 2 for details). This is because the DNA
copy number changes of one gene often influence its own expression
level, and we are less interested in this kind of cis-regulatory
relationships (\texttt{cis-edges}) here.
Furthermore, based on \textit{Exp.Net.664}, no penalties are
imposed on
the expression levels of connected genes either. In another word, we
view the cis-regulations between CNAIs and their linked expression
levels, as well as the inferred RNA interaction network as ``prior
knowledge" in our study.

Note that, different response variables (gene expressions) now have
different sets of predictors, as their neighborhoods in Exp.Net.664
are different. However, the \texttt{remMap} model can still be
fitted with a slight
 modification. The idea is to treat all CNAI ($384$ in total),
 all gene expressions
($654$ in total), as well as five subtype
 indicators as nominal predictors. Then, for each target gene, we force the
 coefficients of those gene expressions
 that do not link to it in \textit{Exp.Net.664} to be zero. We can easily
 achieve this by setting
 those coefficients to zero without updating them throughout
 the iterative fitting procedure.

We select tuning parameters $(\lambda_1,\lambda_2)$ in the
\texttt{remMap} model through a 10-fold cross validation as
described in Section~\ref{sec:tuning}. The optimal $(\lambda_1,
\lambda_2$) corresponding to the smallest CV score from a grid
search is $(355.1, 266.7)$. The resulting model contains $56$
trans-regulations in total. In order to further control false
positive findings, we apply the \texttt{cv.vote} procedure with
$V_a=5$, and filter away $13$ out of these $56$ \texttt{trans-edges}
which have not been consistently selected across different CV folds.
The remaining $43$ \texttt{trans-edges} correspond to three
contiguous CNAIs on chromosome 17 and $31$ distinct (unlinked) RNAs.
Figure~\ref{Fig:remMapNet} illustrates the topology of the estimated
regulatory relationships. The detailed annotations of the three
CNAIs and $31$ RNAs are provided in Table~\ref{table:RA:3CNAI} and
Table~\ref{table:RA:RNAs}.
\begin{table}[h]
 \centering \caption{{Genome locations of
the three CNAIs having (estimated) trans-regulations.}}
\begin{tabular}{c|ccccc}
\hline\hline Index & Cytoband & Begin$^1$ & End $^1$ & $\#$ of
clones$^2$ & $\#$ of Trans-Reg$^3$\\\hline
1 & 17q12-17q12 & 34811630  &  34811630  &  1  & 12 \\
2 & 17q12-17q12 & 34944071  &  35154416  &  9  & 30\\
3 & 17q21.1-17q21.2 &35493689 & 35699243 &  7  &  1\\
\hline\hline \multicolumn{6}{l}{{\footnotesize 1. Nucleotide position (bp). }}\\
\multicolumn{6}{l}{{\footnotesize 2. Number of genes/clones on the array falling into the CNAI. }}\\
\multicolumn{6}{l}{{\footnotesize 3. Number of unlinked genes whose expressions are estimated to be regulated by the CNAI. }}\\
\end{tabular}\label{table:RA:3CNAI}
\end{table}
Moreover, the Pearson-correlations between the DNA copy numbers of
CNAIs and the expression levels of the regulated genes/clones
(including both \texttt{cis-regulation} and
\texttt{trans-regulation}) across the $172$ samples are reported in
Table~\ref{table:RA:RNAs}. As expected, all the cis-regulations have
much higher correlations than the potential trans-regulations. In
addition, none of the subtype indicator variables is selected into
the final model, which implies that the detected associations
between copy numbers of CNAIs and gene expressions are unlikely due
to the stratification of the five tumor subtypes.

\begin{table}[h]
 \centering \caption{{RNAs$^{1}$ being
influenced by the amplifications of the three CNAIs in
Table~\ref{table:RA:3CNAI}. }}
\begin{tabular}{lccr}
\hline\hline Clone ID & Gene symbol & Cytoband & Correlation \\\hline
753692&ABLIM1&10q25&0.199 \\
896962&ACADS&12q22-qter&-0.22 \\
753400&ACTL6A&3q26.33&0.155 \\
472185&ADAMTS1&21q21.2&0.214 \\
210687&AGTR1&3q21-q25&-0.182 \\
856519&ALDH3A2&17p11.2&-0.244 \\
270535&BM466581&19&0.03 \\
238907&CABC1&1q42.13&-0.174 \\
773301&CDH3&16q22.1&0.118 \\
505576&CORIN&4p13-p12&0.196 \\
223350&CP&3q23-q25&0.184 \\
810463&DHRS7B&17p12&-0.151 \\
50582&FLJ25076&5p15.31&0.086 \\
669443&HSF2&6q22.31&0.207 \\
743220&JMJD4&1q42.13&-0.19 \\
43977&KIAA0182&16q24.1&0.259 \\
810891&LAMA5&20q13.2-q13.3&0.269 \\
247230&MARVELD2&5q13.2&-0.214 \\
812088&NLN&5q12.3&0.093 \\
257197&NRBF2&10q21.2&0.275 \\
782449&PCBP2&12q13.12-q13.13&-0.079 \\
796398&PEG3&19q13.4&0.169 \\
293950&PIP5K1A&1q22-q24&-0.242 \\
128302&PTMS&12p13&-0.248 \\
146123&PTPRK&6q22.2-q22.3&0.218 \\
811066&RNF41&12q13.2&-0.247 \\
773344&SLC16A2&Xq13.2&0.24 \\
1031045&SLC4A3&2q36&0.179 \\
141972&STT3A&11q23.3&0.182 \\
454083&TMPO&12q22&0.175 \\
825451&USO1&4q21.1&0.204 \\\hline
68400&BM455010&17&0.748 \\
756253,365147 &ERBB2&17q11.2-q12|17q21.1&0.589 \\
510318,236059 &GRB7&17q12&0.675 \\
245198&MED24&17q21.1&0.367 \\
825577&STARD3&17q11-q12&0.664 \\
782756$^{2}$&TBPL1&6q22.1-q22.3&0.658 \\
\hline\hline \multicolumn{4}{l}{{\footnotesize 1. The first part of
the table lists the inferred trans-regulated genes. The second }}\\
\multicolumn{4}{l}{{\footnotesize part of the
table lists cis-regulated genes.}}\\
\multicolumn{4}{l}{{\footnotesize 2. This cDNA sequence probe is
annotated with \texttt{TBPL1}, but actually }}\\
\multicolumn{4}{l}{{\footnotesize maps to one  of the 17q21.2
genes.}}\\
\end{tabular}\label{table:RA:RNAs}
\end{table}

The three CNAIs being identified as trans-regulators sit closely on
chromosome $17$, spanning from 34811630\textit{bp} to
35699243\textit{bp} and falling into cytoband 17q12-q21.2. This
region (referred to as CNAI-17q12 hereafter) contains 24 known
genes, including the famous breast cancer oncogene ERBB2, and the
growth factor receptor-bound protein 7 (GRB7). The over expression
of GRB7 plays pivotal roles in activating signal transduction and
promoting tumor growth in breast cancer cells with chromosome
17q11-21 amplification~\shortcite{Bai:2008}. In this study,
CNAI-17q12 is highly amplified (normalized $log_2$ ratio$>5$) in 33
($19\%$) out of the $172$ tumor samples. Among the $654$
genes/clones considered in the above analysis, $8$ clones
(corresponding to six genes including ERBB2, GRB7, and MED24) fall
into this region. The expressions of these 8 clones are all
up-regulated by the amplification of CNAI-17q12 (see
Table~\ref{table:RA:RNAs} for more details), which is consistent
with results reported in the literature~\shortcite{Kao:2006}. More
importantly, as suggested by the result of the \texttt{remMap}
model, the amplification of CNAI-17q12 also influences the
expression levels of 31 unlinked genes/clones. This implies that
CNAI-17q12 may harbor transcriptional factors whose activities
closely relate to breast cancer. Indeed, there are 4 transcription
factors (NEUROD2, IKZF3, THRA, NR1D1) and 2 transcriptional
co-activators (MED1, MED24) in CNAI-17q12. It is possible that the
amplification of CNAI-17q12 results in the over expression of one or
more transcription factors/co-activators in this region, which then
influence the expressions of the unlinked 31 genes/clones. In
addition, some of the 31 genes/clones have been reported to have
functions directly related to cancer and may serve as potential drug
targets (see Appendix C.6 of the supplementary material for more
details). In the end, we want to point out that, besides RNA
interactions and subtype stratification, there could be other
unaccounted confounding factors. Therefore, caution must be applied
when one tries to interpret these results.

%% file: Discussion_042309.tex
\section{Discussion}
In this paper, we propose the \texttt{remMap} method for fitting
multivariate regression models under the large $P,Q$ setting. We
focus on model selection, i.e., the  identification of relevant
predictors for each response variable. \texttt{remMap} is motivated
by the rising needs to investigate the regulatory relationships
between different biological molecules based on multiple types of
high dimensional omics data. Such genetic regulatory networks are
usually intrinsically sparse and harbor hub structures. Identifying
the hub regulators (master regulators) is of particular interest, as
they play crucial roles in shaping network functionality. To tackle
these challenges, \texttt{remMap} utilizes a \texttt{MAP} penalty,
which consists of an $\ell_1$ norm part for controlling the overall
sparsity of the network, and an $\ell_2$ norm part for further
imposing a row-sparsity of the coefficient matrix, which facilitates
the detection of master predictors (regulators). This combined
regularization takes into account both model interpretability and
computational tractability. Since the \texttt{MAP} penalty is
imposed on the coefficient matrix as a whole, it helps to borrow
information across different regressions and thus incorporates the
relationships among response variables into the model. As
illustrated in Section 3, this type of ``joint" modeling greatly
improves model efficiency. Also, the combined $\ell_1$ and $\ell_2$
norm penalty further enhances the performance on both edge detection
and master predictor identification.

We also propose a \texttt{cv.vote} procedure to make better use of
the cross validation results. The idea is to treat the training data
from each cross-validation fold as a ``bootstrap" sample and
identify the variables consistently selected in the majority of the
cross-validation folds. As suggested by the simulation study, this
procedure is very effective in decreasing the number of false
positives while only slightly increases the number of false
negatives. Moreover, \texttt{cv.vote} can be applied to a broad
range of model selection problems when cross validation is employed.


In the real application, we apply the \texttt{remMap} method on a
breast cancer data set. Our goal is  to investigate the influences
of DNA copy number alterations on RNA transcript levels based on
$172$ breast cancer tumor samples. The resulting model suggests the
existence of a trans-hub region on cytoband 17q12-q21, whose
amplification influences RNA levels of $31$ unlinked genes. Cytoband
17q12-q21 is a well known hot region for breast cancer, which
harbors the oncogene ERBB2. The above results suggest that this
region may also harbor important transcriptional factors. While our
findings are intriguing, clearly additional investigation is
warranted. One way to verify the above conjecture is through a
sequence analysis to search for common motifs in the upstream
regions of the 31 RNA transcripts, which remains as our future work.

Besides the above application, the \texttt{remMap} model can be
applied to investigate the regulatory relationships between other
types of biological molecules. For example, it is of great interest
to understand the influence of single nucleotide polymorphism (SNP)
on RNA transcript levels, as well as the influence of RNA transcript
levels on protein expression levels. Such investigation will improve
our understanding of related biological systems as well as disease
pathology. In addition, we can utilize the \texttt{remMap} idea to
other models. For example, when selecting a group of variables in a
multiple regression model, we can impose both the  $\ell_2$ penalty
(that is,  the \texttt{group lasso} penalty), as well as an $\ell_1$
penalty to encourage within group sparsity. Similarly, the
\texttt{remMap} idea can also be applied to vector autoregressive
models and generalize linear models.

R package \texttt{remMap} is public available through CRAN ({\small
$http://cran.r-project.org/$}).

\section*{Acknowledgement}
We are grateful to two anonymous reviewers for their valuable
comments.

%% file: Figure_032409.tex

\begin{figure}[h]
\begin{center}
\subfigure[Impact of signal size $s$. $P=Q=600$, $N=200$;
$\rho_x=0.4$; $\rho_\varepsilon=0$; the total number of
\texttt{trans-edges} is $132$.] { \label{Fig:signal}
\includegraphics[width=5.8in, angle=0]{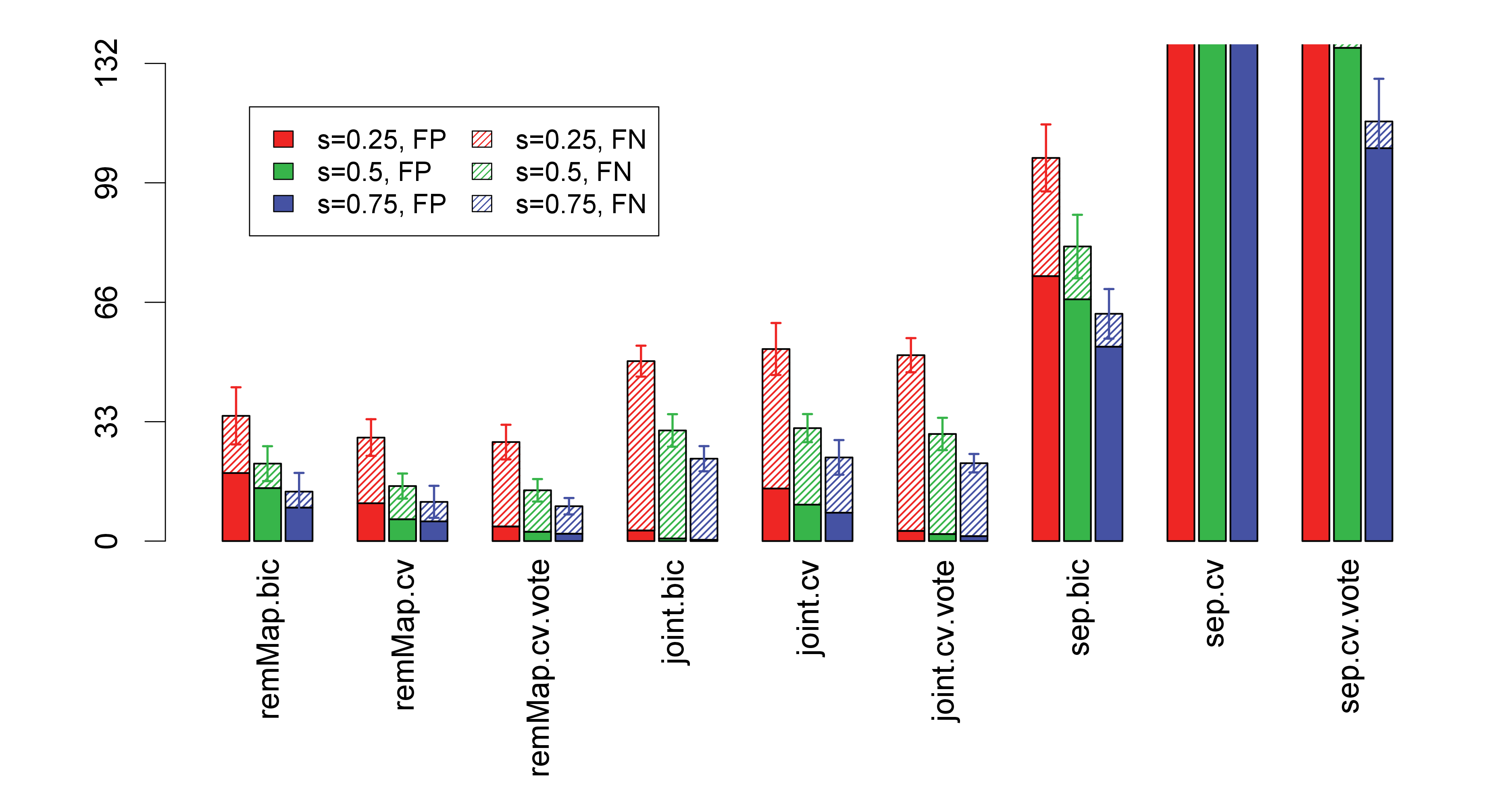}
} \subfigure[Impact of predictor and response dimensionality $P$
$(Q=P)$. $N=200$; $s=0.25$; $\rho_x=0.4$; $\rho_\varepsilon=0$; the
total number of \texttt{trans-edges} is $132$.]{ \label{Fig:dim}
\includegraphics[width=5.8in,  angle=0]{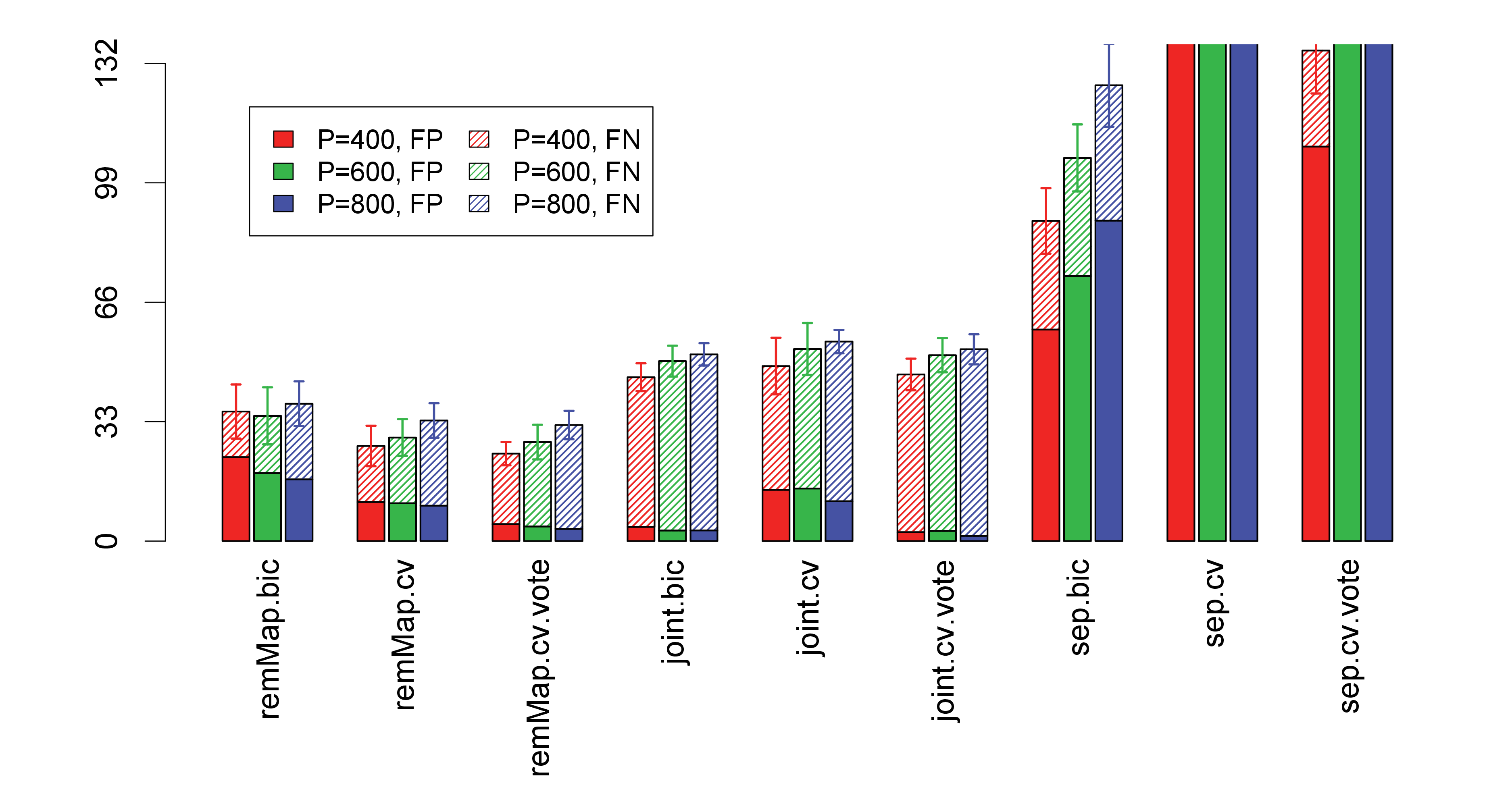}
}
\caption{\small{Impact of signal size and dimensionality. Heights
    of solid bars represent numbers of false positive
    detections of \texttt{trans-edges} (FP); heights of shaded
    bars represent numbers of false negative detections of
    \texttt{trans-edges} (FN). All bars are truncated at
    height=$132$.}} \label{Fig:1}
\end{center}
\end{figure}

\begin{figure}[h]
\begin{center}
\subfigure[Impact of predictor correlation $\rho_x$. $P=Q=600$,
$N=200$; $s=0.25$; $\rho_\varepsilon=0$; the total number of
\texttt{trans-edges} is $132$.] { \label{Fig:corr}
\includegraphics[width=5.8in, angle=0]{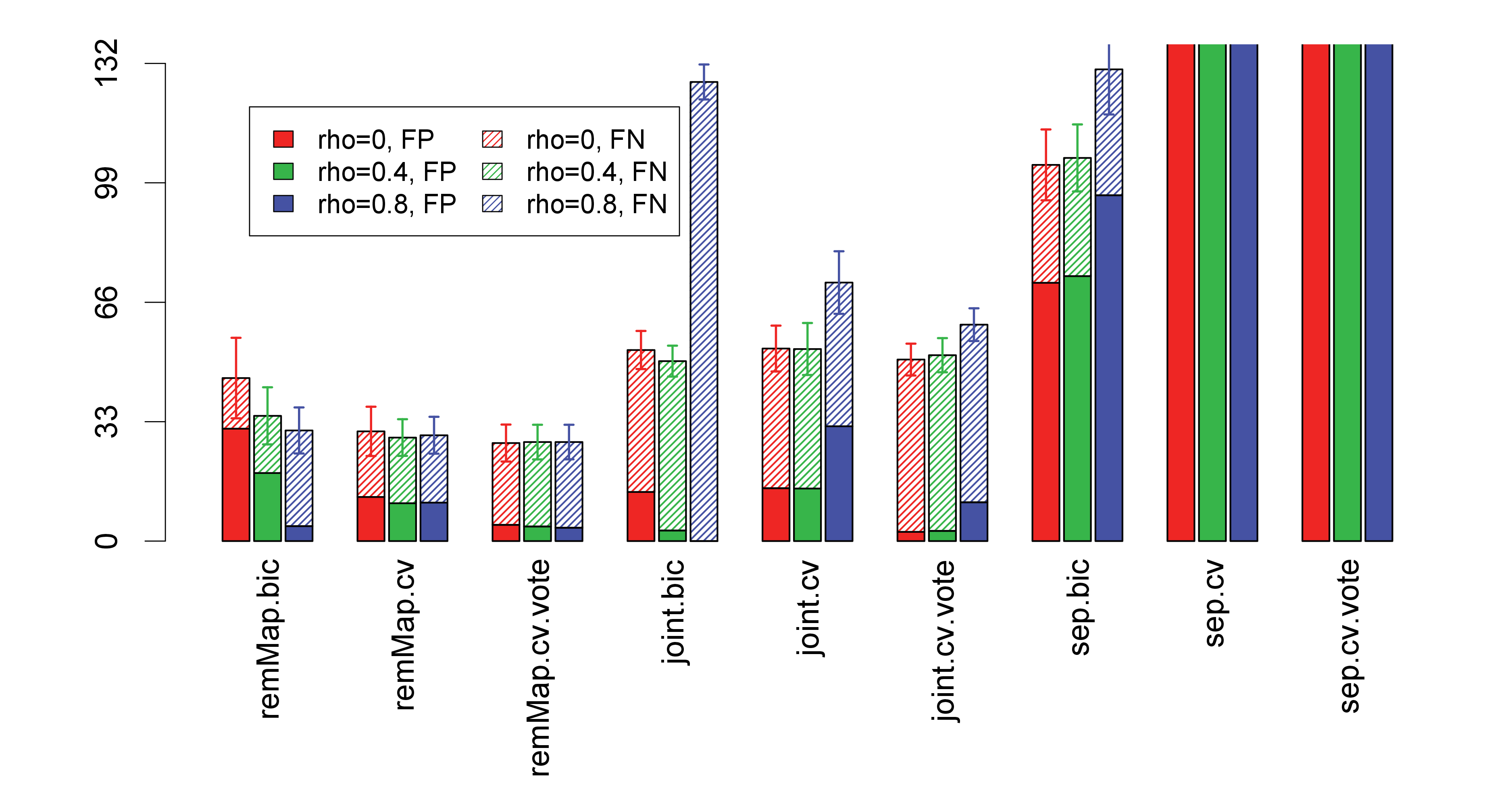}
} \subfigure[Impact of residual correlation $\rho_\varepsilon$.
$P=Q=600$, $N=200$;
  $s=0.25$; $\rho_x=0.4$; the total number of \texttt{trans-edges} is
  $132$.]
{\label{Fig:corr_res}
  \includegraphics[width=5.8in, angle=0]{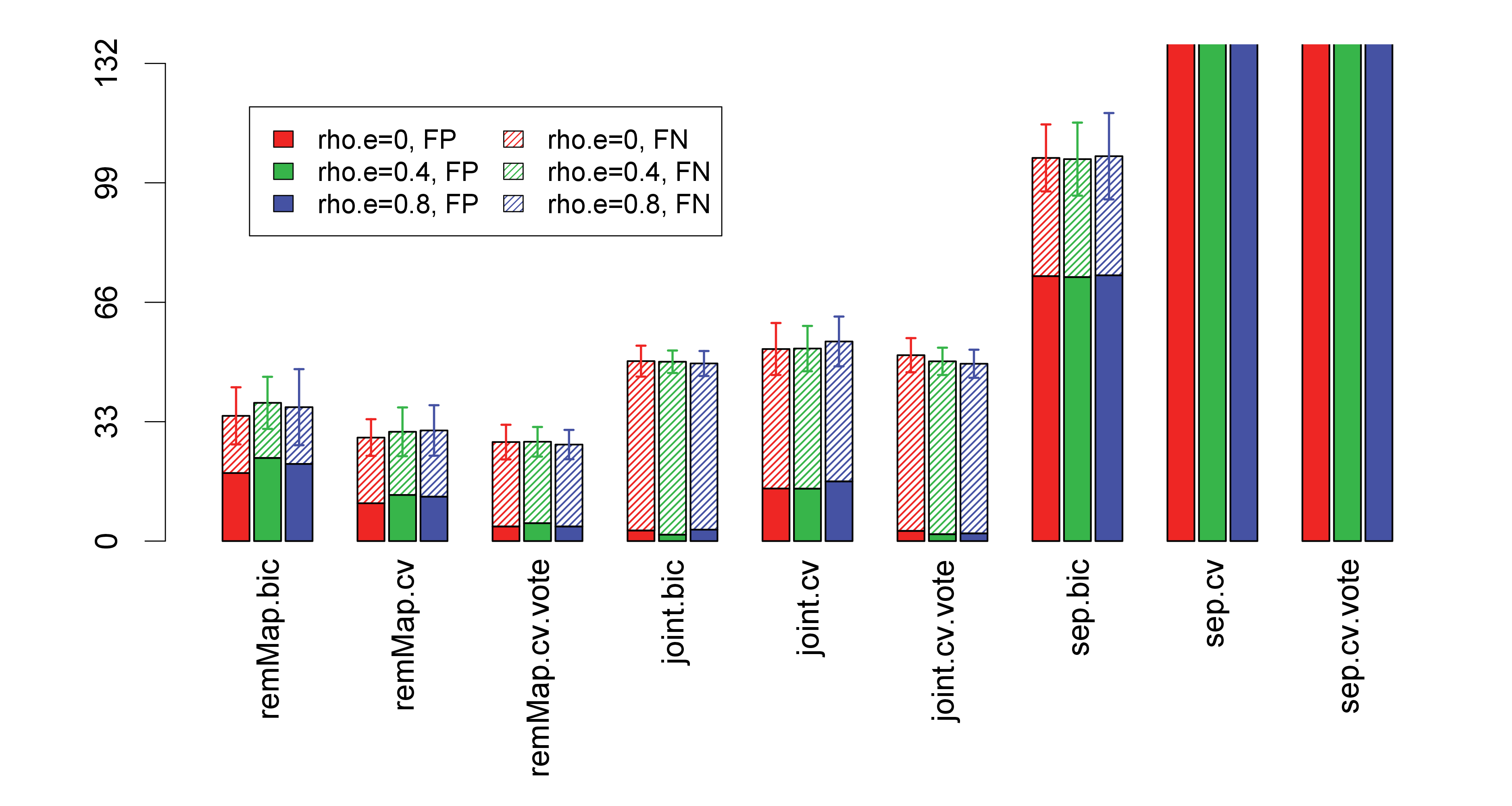}}
\caption{\small{Impact of correlations. Heights of solid bars
    represent numbers of false positive detections of
    \texttt{trans-edges} (FP); heights of shaded bars
    represent numbers
    of false negative detections of \texttt{trans-edges}
    (FN). All bars are truncated at height=$132$. }} \label{Fig:2}
\end{center}
\end{figure}


\begin{figure}[h]
\begin{center}
\includegraphics[width=3in, height=1in, angle=0]{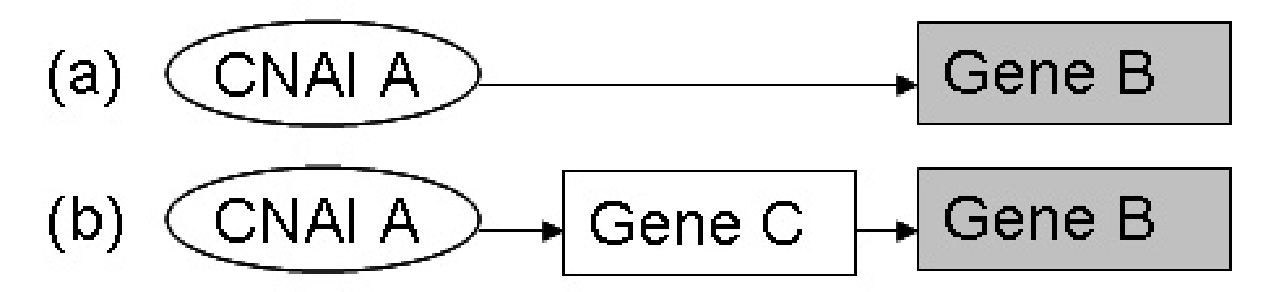}
\caption{(a) Direct interaction between CNAI A and the expression of
gene B; (b) indirect interaction between CNAI A and the expression
of Gene B through one intermediate gene.} \label{figure:diagram}
\end{center}
\end{figure}

\begin{figure}[h]
\begin{center}
\includegraphics[width=11cm, angle=0]{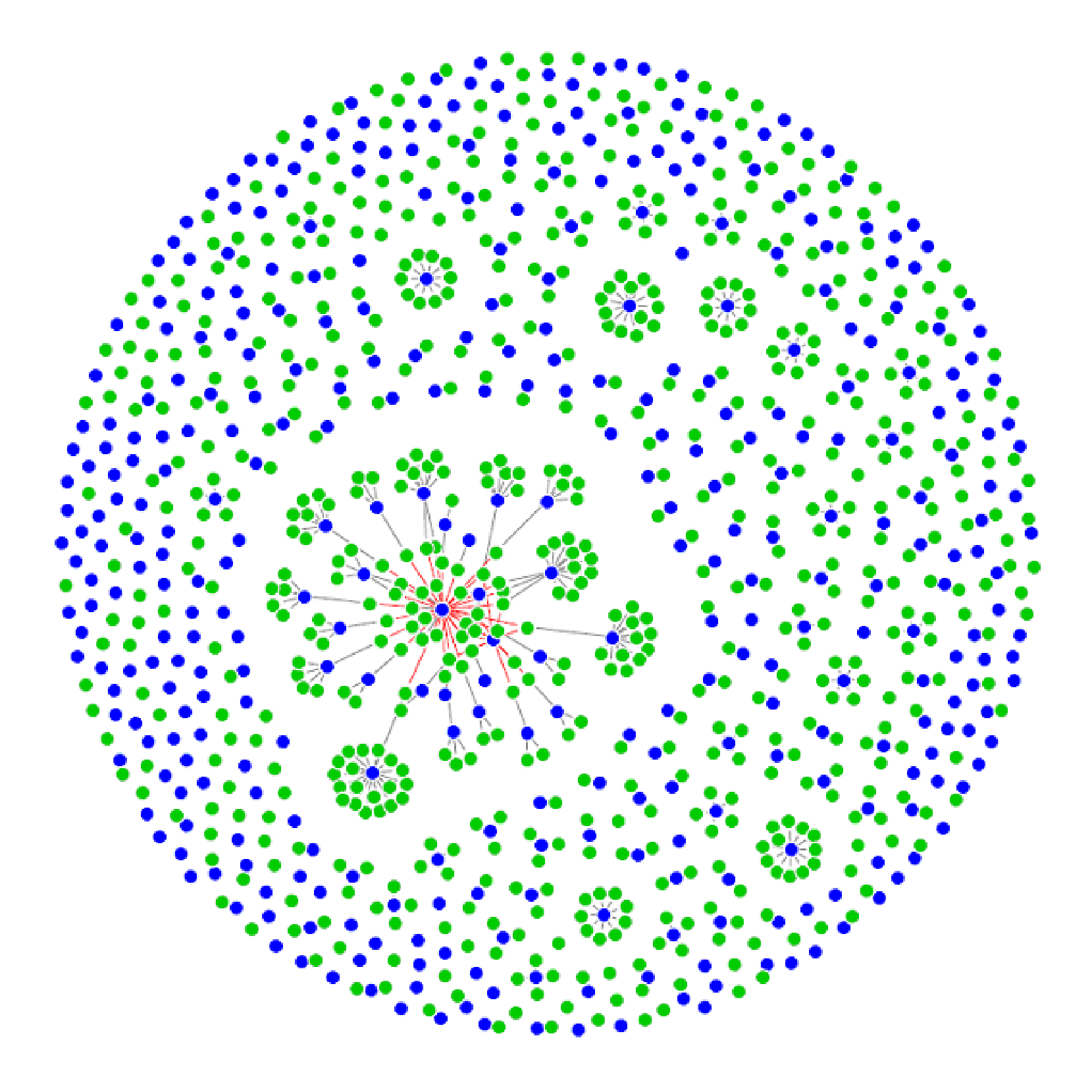}
\caption{Network of  the estimated regulatory relationships between
the copy numbers of the 384 CNAIs and the expressions of the 654
breast cancer related genes. Each blue node stands for one CNAI, and
each green node stands for one gene. Red edges
represent inferred trans-regulations (43 in total). Grey edges
represent cis-regulations.}\label{Fig:remMapNet}
\end{center}
\vspace{-23pt}
\end{figure}

%% file: AppendixAB_042309.tex
{\centering {\Large \bf Supplementary Material}}

\renewcommand{\theequation}{S-\arabic{equation}}

\section*{Appendix A: Proof of Theorem \ref{thm:fitting}}\label{sec:AppendixA}
 Define
\begin{eqnarray*}
L(\beta;Y,X)=\frac{1}{2}\sum_{q=1}^Q (y_q-x\beta_q)^2+\lambda_1
\sum_{q=1}^Q |\beta_q|+\lambda_2 \sqrt{\sum_{q=1}^Q \beta_q^2}.
\end{eqnarray*}
It is obvious that, in order to prove Theorem \ref{thm:fitting}, we
only need to show that, the solution of $\min_{\beta} L(\beta;Y,X)$,
is given by (for $q=1,\cdots,Q$)
\begin{eqnarray*}
\widehat{\beta}_q=\left\{ \begin{array}{ll} 0, &\textrm{if}~
||\widehat{\beta}^{\textrm{lasso}}||_2=0;\\
\widehat{\beta}^{\textrm{lasso}}_q \left(1 -
\frac{\lambda_2}{||\widehat{\beta}^{\textrm{lasso}}||_2
x^2}\right)_{+}, &\textrm{otherwise},\\
\end{array}\right.
\end{eqnarray*}
where
\begin{eqnarray}
\label{eqn:lasso_one}
\widehat{\beta}^{\textrm{lasso}}_q=\left(1-\frac{\lambda_1}{|xy_q|}\right)_{+}\frac{xy_q}{x^2}.
\end{eqnarray}
In the following, for function $L$, view $\{\beta_{q^{\prime}}:
q^{\prime} \neq q\}$ as fixed. With a slight abuse of notation,
write $L=L(\beta_q)$. Then when $\beta_q \geq 0$, we have
$$
\frac{d L}{d
\beta_q}=-xy_q+(x^2+\frac{\lambda_2}{||\beta||_2})\beta_q+\lambda_1.
$$
Thus, $\frac{d L}{d \beta_q}>0$ if and only if $\beta_q>
\tilde{\beta_q}^+$, where
$$
\tilde{\beta_q}^+ :=
\frac{xy_q}{x^2+\frac{\lambda_2}{||\beta||_2}}(1-\frac{\lambda_1}{xy_q}).
$$
Denote the minima of $L(\beta_q)|_{\beta_q \geq 0}$ by
$\beta_{q,\textrm{min}}^+$. Then, when $\tilde{\beta_q}^+>0$,
$\beta_{q,\textrm{min}}^+=\tilde{\beta_q}^+$. On the other hand,
when $\tilde{\beta_q}^+\leq 0$, $\beta_{q,\textrm{min}}^+=0$. Note
that $\tilde{\beta_q}^+ >0$ if and only if
$xy_q(1-\frac{\lambda_1}{xy_q})>0$. Thus we have
\begin{eqnarray*}
\beta_{q,\textrm{min}}^+=\left\{\begin{array}{ll} \tilde{\beta}_q^+,
&\textrm{if}~
xy_q(1-\frac{\lambda_1}{xy_q})>0;\\
0, &\textrm{if}~ xy_q(1-\frac{\lambda_1}{xy_q})\leq0.\\
\end{array}\right.
\end{eqnarray*}
Similarly, denote the minima of $L(\beta_q)|_{\beta_q \leq 0}$ by
$\beta_{q,\textrm{min}}^-$, and define
$$
\tilde{\beta_q}^- :=
\frac{xy_q}{x^2+\frac{\lambda_2}{||\beta||_2}}(1+\frac{\lambda_1}{xy_q}).
$$
Then we have
\begin{eqnarray*}
\beta_{q,\textrm{min}}^-=\left\{\begin{array}{ll} \tilde{\beta}_q^-,
&\textrm{if}~
xy_q(1+\frac{\lambda_1}{xy_q})<0;\\
0, &\textrm{if}~ xy_q(1+\frac{\lambda_1}{xy_q})\geq 0.\\
\end{array}\right.
\end{eqnarray*}
Denote the minima of $L(\beta_q)$ as $\widehat{\beta}_q$ (with a
slight abuse of notation). From the above, it is obvious that, if
$xy_q>0$, then $\widehat{\beta}_q \geq 0$. Thus
$\widehat{\beta}_q=\max(\tilde{\beta}_q^+,0)=\frac{xy_q}{x^2+\frac{\lambda_2}{||\beta||_2}}
(1-\frac{\lambda_1}{xy_q})_{+}=\frac{xy_q}{x^2+\frac{\lambda_2}{||\beta||_2}}(1-\frac{\lambda_1}{|xy_q|})_{+}$.
Similarly, if $xy_q \leq 0$, then $\widehat{\beta}_q \leq 0$, and it
has the same expression as above. Denote the minima of
$L(\beta)|_{||\beta||_2 >0}$ (now viewed as a function of
$(\beta_1,\cdots,\beta_Q)$) as
$\widehat{\beta}_{\textrm{min}}=(\widehat{\beta}_{1,\textrm{min}},\cdots,\widehat{\beta}_{Q,\textrm{min}})$.
We have shown above that, if such a minima exists, it satisfies (for
$q=1,\cdots,Q$)
\begin{eqnarray}
\label{eqn:beta_min}
\widehat{\beta}_{q,\textrm{min}}=\frac{xy_q}{x^2+\frac{\lambda_2}{||\widehat{\beta}_{\textrm{min}}||_2}}\left(1-\frac{\lambda_1}{|xy_q|}\right)_{+}
=\widehat{\beta}_q^{\textrm{lasso}}\frac{x^2}{x^2+\frac{\lambda_2}{||\widehat{\beta}_{\textrm{min}}||_2}},
\end{eqnarray}
where $\widehat{\beta}_q^{\textrm{lasso}}$ is defined by equation
(\ref{eqn:lasso_one}). Thus
\begin{eqnarray*}
||\widehat{\beta}_{\textrm{min}}||_2=||\widehat{\beta}^{\textrm{lasso}}||_2\frac{x^2}{x^2+\frac{\lambda_2}{||\widehat{\beta}_{\textrm{min}}||_2}}.
\end{eqnarray*}
By solving the above equation, we obtain
$$
||\widehat{\beta}_{\textrm{min}}||_2=||\widehat{\beta}^{\textrm{lasso}}||_2-\frac{\lambda_2}{x^2}.
$$
By plugging the expression on the right hand side into
(\ref{eqn:beta_min}), we achieve
$$
\widehat{\beta}_{q,\textrm{min}}=\widehat{\beta}_q^{\textrm{lasso}}
\left(1-\frac{\lambda_2}{||\widehat{\beta}^{\textrm{lasso}}||_2x^2}\right).
$$
Denote the minima of $L(\beta)$ by
$\widehat{\beta}=(\widehat{\beta}_1,\cdots,\widehat{\beta}_Q)$. From
the above, we also know that  if
$||\widehat{\beta}^{\textrm{lasso}}||_2-\frac{\lambda_2}{x^2}>0$,
$L(\beta)$ achieves its minimum on $||\beta||_2>0$, which is
$\widehat{\beta}=\widehat{\beta}_{\textrm{min}}$. Otherwise,
$L(\beta)$ achieves its minimum at zero. Since
$||\widehat{\beta}^{\textrm{lasso}}||_2-\frac{\lambda_2}{x^2}>0$ if
and only if
$1-\frac{\lambda_2}{||\widehat{\beta}^{\textrm{lasso}}||_2x^2}>0$,
we have proved the theorem.

\section*{Appendix B: BIC criterion for tuning}\label{sec:AppendixB}


In this section, we describe the BIC criterion for selecting
$(\lambda_1,\lambda_2)$. We also derive an unbiased estimator of the
degrees of freedom of the \texttt{remMap} estimator under orthogonal
design.

In model (\ref{eqn:mr}), by assuming $\epsilon_q \sim {\rm
Normal}(0,\sigma^2_{q,\epsilon})$, the BIC criterion for the $q^{th}$ regression can be
defined as
\begin{eqnarray}
\label{eqn:bic_q} {\rm
BIC}_q(\widehat{\beta}_{1q},\cdots,\widehat{\beta}_{Pq};{\rm
df}_q)=N \times \log({\rm RSS}_q)+\log N \times {\rm df}_q ,
\end{eqnarray}
where ${\rm RSS}_q:=\sum_{n=1}^N(y^n_q-\widehat{y}^n_q)^2$ with
$\widehat{y}^n_q=\sum_{p=1}^P x^n_p \widehat{\beta}_{pq}$; and
${\rm df}_q$ is the degrees of freedom which is defined as
\begin{eqnarray}
\label{eqn:df} {\rm df}_q={\rm
df}_q(\widehat{\beta}_{1q},\cdots,\widehat{\beta}_{Pq}):=\sum_{n=1}^N{\rm
Cov}(\widehat{y}^n_q,y^n_q)/\sigma^2_{q,\epsilon},
\end{eqnarray}
 where $\sigma^2_{q,\epsilon}$ is the variance of
$\epsilon_q$.

For a given pair of $(\lambda_1,\lambda_2)$,
We then define the (overall) BIC criterion at $(\lambda_1,\lambda_2)$:
\begin{eqnarray}
\label{eqn:bic} {\rm BIC}(\lambda_1,\lambda_2)
=N\times \sum_{q=1}^Q\log({\rm RSS}_q(\lambda_1,\lambda_2))+\log N
\times \sum_{q=1}^Q{\rm df}_q(\lambda_1,\lambda_2).
\end{eqnarray}

\shortciteN{lars} derive an explicit formula for the degrees of
freedom of \textit{lars} under orthogonal design. Similar strategy
are also used by \shortciteN{YuanLin:2006} among others. In the
following theorem, we follow the same idea and derive an unbiased
estimator of $df_q$ for \texttt{remMap} when the columns of
$\mathbf{X}$ are orthogonal to each other.

{\thm \label{thm:df}Suppose $X_p^T X_{p^{\prime}}=0$ for all $1 \leq
p \not= p^{\prime} \leq P$. Then for given $(\lambda_1,\lambda_2)$,
\begin{eqnarray}\label{eqn:df_rem_q}
\widehat{df}_q(\lambda_1,\lambda_2)&:=&\sum_{p=1}^P c_{pq}\times
\mathbb{I} \left(||\widehat{B}_p^{\rm
lasso}||_{2,C}>\frac{\lambda_2}{||X_p||^2_2}\right)\times
\mathbb{I} \left(|\widehat{\beta}_{pq}^{\rm
ols}|>\frac{\lambda_1}{||X_p||^2_2}\right)\nonumber \\
&\times& \left(1-\frac{\lambda_2}{||X_p||^2_2}
\frac{||\widehat{B}_p^{\rm lasso}||_{2,C}^2-(\widehat{\beta}^{\rm
lasso}_{pq})^2}{||\widehat{B}_p^{\rm
lasso}||_{2,C}^3}\right)+\sum_{p=1}^P(1-c_{p,q})
\end{eqnarray}
is an unbiased estimator of the degrees of freedom  ${\rm
df}_q(\lambda_1,\lambda_2)$ (defined in equation (\ref{eqn:df})) of
the \texttt{remMap} estimator
$\widehat{\mathbf{B}}=\widehat{\mathbf{B}}(\lambda_1,\lambda_2)=(\widehat{\beta}_{pq}(\lambda_1,\lambda_2))$.
Here, under the orthogonal design, $\widehat{\beta}_{pq},
\widehat{\beta}_{pq}^{\rm lasso}$ are given by Theorem
\ref{thm:fitting} with $\widetilde{Y}_q=Y_q$ ($q=1,\cdots,Q$), and
$\widehat{\beta}^{\rm ols}_{pq}:=\frac{X_p^T Y_q}{||X_p||^2_2}$. }

Before proving Theorem \ref{thm:df}, we first explain definition
(\ref{eqn:df}) -- the degrees of freedom. Consider the $q^{th}$
regression in model (\ref{eqn:mr}). Suppose that
$\{\widehat{y}^n_q\}_{n=1}^N$ are the fitted values by a certain
fitting procedure  based on the current observations
$\{y^n_q: n=1,\cdots,N; q=1,\cdots,Q\}$. Let
$\mu^n_q:=\sum_{p=1}^P x^n_p \beta_{pq}$. Then for a fixed design
matrix $X=(x^n_p)$,  the expected re-scaled prediction error of
$\{\widehat{y}^n_q\}_{n=1}^N$ in predicting a future set of new
observations $\{\widetilde{y}^n_q\}_{n=1}^N$ from the $q^{th}$
regression of model (\ref{eqn:mr}) is:
$$
{\rm PE}_q=\sum_{n=1}^N
E((\widehat{y}^n_q-\widetilde{y}^n_q)^2)/\sigma^2_{q,\epsilon}=\sum_{n=1}^N
E((\widehat{y}^n_q-\mu^n_q)^2)/\sigma^2_{q,\epsilon}+N.
$$

Note that
$$
(\widehat{y}^n_q-y^n_q)^2=(\widehat{y}^n_q-\mu^n_q)^2+(y^n_q-\mu^n_q)^2-2(\widehat{y}^n_q-\mu^n_q)(y^n_q-\mu^n_q).
$$
Therefore,
$$
{\rm PE}_q=\sum_{n=1}^N
E((\widehat{y}^n_q-y^n_q)^2)/\sigma^2_{q,\epsilon}+2\sum_{n=1}^N
{\rm Cov}(\widehat{y}^n_q,y^n_q)/\sigma^2_{q,\epsilon}.
$$
Denote  $RSS_q=\sum_{n=1}^N (\widehat{y}^n_q-y^n_q)^2$. Then an
un-biased estimator of ${\rm PE}_q$ is
$$
RSS_q/\sigma^2_{q,\epsilon}+2\sum_{n=1}^N {\rm
Cov}(\widehat{y}^n_q,y^n_q)/\sigma^2_{q,\epsilon}.
$$
Therefore, a natural definition of the degrees of freedom for the
procedure resulting the fitted values $\{\widehat{y}^n_q\}_{n=1}^N$
is as given in equation  (\ref{eqn:df}). Note that, this is the
definition used in Mallow's $C_p$ criterion.

\medskip
 \noindent{\textbf{Proof of Theorem \ref{thm:df}}}: By applying Stein's identity to the Normal distribution,  we have: if $Z \sim N(\mu,\sigma^2)$, and a function $g$ such that $E(|g^\prime(Z)|)<\infty$, then
 $$
 {\rm Cov}(g(Z),Z)/\sigma^2=E(g^\prime(Z)).
 $$
 Therefore, under the normality assumption on the residuals $\{\epsilon_q\}_{q=1}^Q$ in model (\ref{eqn:mr}), definition (\ref{eqn:df}) becomes
 $$
 df_q=\sum_{n=1}^N E\left(\frac{\partial \widehat{y}^n_q}{\partial y^n_q} \right), \quad q=1,\cdots, Q.
 $$
Thus an obvious unbiased estimator of $df_q$ is
$\widehat{df}_q=\sum_{n=1}^N \frac{\partial
\widehat{y}^n_q}{\partial y^n_q}$. In the following, we derive
$\widehat{df}_q$ for the proposed \texttt{remMap} estimator under
the orthogonal design. Let
$\widehat{\beta}_q=(\widehat{\beta}_{1q},\cdots,\widehat{\beta}_{Pq})$
be a one by $P$ row vector; let $\mathbf{X}=(x^n_p)$ be the $N$ by
$P$ design matrix which is orthonormal; let
$Y_q=(y^1_1,\cdots,y^N_q)^T$ and
$\widehat{Y}_q=(\widehat{y}^1_q,\cdots,\widehat{y}^N_q)^T=\mathbf{X}\widehat{\beta}_q$
be $N$ by one column vectors. Then
$$
 \widehat{df}_q={\rm tr}\left(\frac{\partial \widehat{Y}_q}{\partial Y_q}\right)={\rm tr}\left( \mathbf{X} \frac{\partial \widehat{\beta}_q}{\partial Y_q}\right)={\rm tr}\left( \mathbf{X}  \frac{\partial \widehat{\beta}_q}{\partial \widehat{\beta}_{q,\textrm{ols}}} \frac{\partial \widehat{\beta}_{q,\textrm{ols}}}{\partial Y_q}\right),
$$
where
$\widehat{\beta}_{q,\textrm{ols}}=(\widehat{\beta}^{\textrm{ols}}_{1q},\cdots,\widehat{\beta}^{\textrm{ols}}_{Pq})^T$
and the last equality is due to the chain rule. Since under the
orthogonal design,
$\widehat{\beta}^{\textrm{ols}}_{pq}=X_p^TY_q/||X_p||^2_2$, where
$X_p=(x^1_p,\cdots,x^N_p)^T$, thus $ \frac{\partial
\widehat{\beta}_{q,\textrm{ols}}}{\partial Y_q}=\mathbf{D} \mathbf{X}^T,$ where $\mathbf{D}$ is a $P$ by
$P$ diagonal matrix with the $p^{th}$ diagonal entry being
$1/||X_p||^2_2$. Therefore
$$
\widehat{df}_q={\rm tr}\left( \mathbf{X}  \frac{\partial
\widehat{\beta}_q}{\partial \widehat{\beta}_{q,\textrm{ols}}}  \mathbf{D} \mathbf{X}^T
\right)={\rm tr}\left( \mathbf{D}  \mathbf{X}^T  \mathbf{X}  \frac{\partial
\widehat{\beta}_q}{\partial \widehat{\beta}_{q,\textrm{ols}}} \right)={\rm
tr}\left( \frac{\partial \widehat{\beta}_q}{\partial
\widehat{\beta}_{q,\textrm{ols}}} \right)=\sum_{p=1}^P\frac{\partial
\widehat{\beta}_{pq}}{\partial \widehat{\beta}^{\textrm{ols}}_{pq}} ,
$$
where the second to last equality is by $ \mathbf{X}^T \mathbf{X}= \mathbf{D}^{-1}$ which is due
to the orthogonality of $\mathbf{X}$. By the chain rule
$$
\frac{\partial \widehat{\beta}_{pq}}{\partial
\widehat{\beta}^{\textrm{ols}}_{pq}}=\frac{\partial
\widehat{\beta}_{pq}}{\partial
\widehat{\beta}^{\textrm{lasso}}_{pq}}\frac{\partial
\widehat{\beta}^{\textrm{lasso}}_{pq}}{\partial \widehat{\beta}^{\textrm{ols}}_{pq}}.
$$
By Theorem 1, under the orthogonal design,
$$
\frac{\partial \widehat{\beta}_{pq}}{\partial
\widehat{\beta}^{\textrm{lasso}}_{pq}}=\mathbb{I} \left(||
\widehat{B}_p^{\textrm{lasso}}||_{2,C}>\frac{\lambda_2}{||X_p||_2^2}\right)
\times \left[1-\frac{\lambda_2}{||X_p||_2^2}\frac{||
\widehat{B}_p^{\textrm{lasso}}||^2_{2,C}-(\widehat{\beta}^{\textrm{lasso}}_{pq})^2}{||
\widehat{B}_p^{\textrm{lasso}}||^3_{2,C}}\right],
$$
and
$$
\frac{\partial \widehat{\beta}^{\textrm{lasso}}_{pq}}{\partial
\widehat{\beta}^{\textrm{ols}}_{pq}}=\left\{ \begin{array}{ll}
1, &\textrm{if}~ c_{p,q}=0;\\
\mathbb{I}
\left(|\widehat{\beta}^{\textrm{ols}}_{pq}|>\frac{\lambda_1}{||X_p||^2_2}\right),
&\textrm{if}~ c_{p,q}=1.
\end{array}\right.
$$
Note that when $c_{p,q}=0$,
$\widehat{\beta}_{pq}=\widehat{\beta}^{\textrm{ols}}_{pq}$, thus
$\frac{\partial \widehat{\beta}_{pq}}{\partial
\widehat{\beta}^{\textrm{ols}}_{pq}}=1$. It is then easy to show that
$\widehat{df}_q$ is as given in equation (\ref{eqn:df_rem_q}).

Note that,  when the $\ell_2$ penalty parameter
  $\lambda_2$ is 0, the model becomes $q$ separate \texttt{lasso}
  regressions with the same penalty parameter $\lambda_1$ and the
  degrees of freedom estimation in equation (\ref{eqn:df_rem_q}) is simply the
  total number of non-zero coefficients in the model (under
  orthogonal design). When
  $\lambda_2$ is nonzero, the degrees of freedom of
  \texttt{remMap} estimator should be smaller than the number of
  non-zero coefficients due to the additional shrinkage induced
  by the $\ell_2$ norm part of the \texttt{MAP} penalty (equation
  (3)). This is reflected by equation (\ref{eqn:df_rem_q}).

%% file: AppendixC_042309.tex
\section*{Appendix C: Data Preprocessing}
\subsection*{C.1 Preprocessing for array CGH data}\label{sec:realAppendix:CGH}
Each array output ($log_2$ ratios) is first standardized to have
median$=0$ and smoothed by \texttt{cghFLasso} \shortcite{cghFLasso2008} for defining
gained/lost regions on the genome. The
noise level of each array is then calculated based on the
measurements from the estimated normal regions (i.e., regions with
estimated copy numbers equal to 2). After that, each smoothed array
is normalized according to its own noise level.

We define {\it copy number alteration intervals (CNAIs)} by using
the Fixed-Order Clustering (\texttt{FOC}) method
~\shortcite{Wang:thesis}. \texttt{FOC} first builds a hierarchical
clustering tree along the genome based on all arrays, and then cuts
the tree at an appropriate height such that genes with similar copy
numbers fall into the same CNAI. \texttt{FOC} is a generalization of
the \texttt{CLAC} (CLuster Along Chromosome) method proposed by
\shortciteN{CLAC:2005}. It differs in two ways from the standard
agglomerative clustering. First, the order of the leaves in the tree
is fixed, which represents the genome order of the genes/clones in
the array. So, only adjacent clusters are joined together when the
tree is generated by a bottom-up approach. Second, the similarity
between two clusters no longer refers to the spatial distance but to
the similarity of the array measurements ($log_2$ ratio) between the
two clusters. By using \texttt{FOC}, the whole genome is divided
into $384$ non-overlapping CNAIs based on all $172$ CGH arrays. This
is illustrated in Figure~\ref{Fig:app:FOC}. In addition, the heatmap
of the (sample) correlations of the CNAIs is given in Figure S-2.

\subsection*{C.2 Selection of breast cancer related genes}\label{sec:realAppendix:exp}
We combine seven published breast cancer gene lists: the intrinsic
gene set~\shortcite{Sorlie:2003}, the Amsterdam 70
gene~\shortcite{vandeVijver:2002}, the wound response gene set
~\shortcite{Chang:2004}, the 76 genes for the metastasis
signature~\shortcite{Wang:2005}, the gene list for calculating the
recurrence score~\shortcite{Paik:2004}, the gene list of the Genomic
Grade Index (GGI)~\shortcite{Sotiriou:2006}, and the PTEN gene
set~\shortcite{Saal:2007}. Among this union set of breast cancer
related genes, $967$ overlap with the genes in the current
expression data set. We further filter away genes with missing
measurements in more than $20\%$ of the samples, and $654$ genes are
left. Among these $654$ selected genes, $449$ are from the intrinsic
gene set~\shortcite{Sorlie:2003}, which are used to derive breast
cancer subtype labels in Appendix C.4.

\subsection*{C.3 Interactions among RNA expressions}
We apply the \texttt{space} (Sparse PArtial Correlation Estimation)
method ~\shortcite{space:2008} to infer the interactions among RNA
levels through identifying non-zero partial correlations.
\texttt{space} assumes the overall sparsity of the partial
correlation matrix and employs sparse regression techniques for
model fitting. As indicated by many experiments that
genetic-regulatory networks have a power-law type degree
distribution with a power parameter between 2 and
3~\shortcite{Newman:2003}, the tuning parameter in \texttt{space} is
chosen such that the resulting network has an estimated power
parameter around $2$ (see Figure S-3(b)
for the corresponding degree distribution). The resulting
(concentration) network has $664$ edges in total, whose topology is
illustrated in Figure S-3(a).
In this network, there are 7 nodes having at least 10 edges. These
hub genes include PLK1, PTTG1, AURKA, ESR1, and GATA3. PLK1 has
important functions in maintaining genome stability via its role in
mitosis. Its over expression is associated with preinvasive in situ
carcinomas of the breast~\shortcite{Rizki:2007}. PTTG1 is observed
to be a proliferation marker in invasive ductal breast
carcinomas~\shortcite{Talvinen:2008}. AURKA encodes a cell
cycle-regulated kinase and is a potential metastasis promoting gene
for breast cancer~\shortcite{Thomassen:2008}. ESR1 encodes an
estrogen receptor, and is a well known key player in breast cancer.
Moreover, it had been reported that GATA3 expression has a strong
association with estrogen receptor in breast
cancer~\shortcite{Voduc:2008}. Detailed annotation of these and
other hub genes are listed in Table~\ref{table:RA:SpaceHubGene}. We
refer to this network as \textit{Exp.Net.664}, and use it to account
for RNA interactions when investigating the regulations between
CNAIs and RNA levels.
\begin{table}[h]
\centering \caption{{Annotations for hub genes (degrees greater than
10) in the inferred RNA interaction network \textit{Exp.Net.664}.}}
\begin{tabular}{c|llll}
\hline\hline CloneID & Gene Name & Symbol & ID & Cytoband\\\hline
744047 & Polo-like kinase 1 (Drosophila)& PLK1 &
5347 & 16p12.1 \\
781089 & Pituitary tumor-transforming 1 &  PTTG1 & 9232 &
5q35.1\\
129865 & Aurora kinase A & AURKA  & 6790   & 20q13.2-q13.3\\
214068 & GATA binding protein 3 &  GATA3 &   2625  & 10p15\\
950690 & Cyclin A2 & CCNA2 & 890 &  4q25-q31 \\
120881 & RAB31, member RAS oncogene family & RAB31 & 11031 &
18p11.3 \\
725321 & Estrogen receptor 1 & ESR1 & 2099 & 6q25.1\\
\hline\hline
\end{tabular}\label{table:RA:SpaceHubGene}
\end{table}

\subsection*{C.4 Breast Cancer Subtypes}
Population stratification due to distinct subtypes could confound
our detection of associations between CNAIs and gene expressions.
For example, if the copy number of CNAI A and expression level of
gene B are both higher in one subtype than in the other subtypes, we
could observe a strong correlation between CNAI A and gene
expression B across the whole population, even when the correlation
within each subtype is rather weak. To account for this potential
confounding factor, we introduce a set of subtype indicator
variables, which is used as additional predictors in the
\texttt{remMap} model. Specifically, we derive subtype labels based
on expression patterns by following the work
of~\shortciteN{Sorlie:2003}. We first normalize the expression
levels of each intrinsic gene ($449$ in total) across the $172$
samples to have mean zero and MAD one. Then we use \texttt{kmeans}
clustering to divide the patients into five distinct groups, which
correspond to the five subtypes suggested
by~\shortciteN{Sorlie:2003}
--- Luminal Subtype A, Luminal Subtype B, ERBB2-overexpressing
Subtype, Basal Subtype and Normal Breast-like Subtype.
Figure~\ref{Fig:subtype} illustrates the expression patterns of
these five subtypes across the $172$ samples. We then define five
dummy variables to represent the subtype information for each tumor
sample, and include them as predictors when fitting the
\texttt{remMap} model.

\subsection*{C.5 Comments on the results of the \texttt{remMap} analysis.}
\texttt{remMap} analysis suggests that the amplification of a
trans-hub region on chromosome 17 influences the RNA expression
levels of $31$ distinct (unlinked) genes. Some of the 31
genes/clones have been reported to have functions directly related
to cancer and may serve as potential drug targets. For example,
AGTR1 is a receptor whose genetic polymorphisms have been reported
to associate with breast cancer risk and is possibly
druggable~\shortcite{Koh:2005}. CDH3 encodes a cell-cell adhesion
glycoprotein and is deemed as a candidate of tumor suppressor gene,
as disturbance of intracellular adhesion is important for invasion
and metastasis of tumor cells~\shortcite{Kremmidiotis:1998}. PEG3 is
a mediator between p53 and Bax in DNA damage-induced neuronal death
~\shortcite{Johnson:2002} and may function as a tumor suppressor
gene~\shortcite{Dowdy:2005}. In a word, these 31 genes may play
functional roles in the pathogenesis of breast cancer and may serve
as additional targets for therapy.

%% file: remMap_Appendix.bbl
{}

%% file: Figure_Appendix_042309.tex
\begin{figure}[h]
\begin{center}
\includegraphics[width=15cm, angle=0]{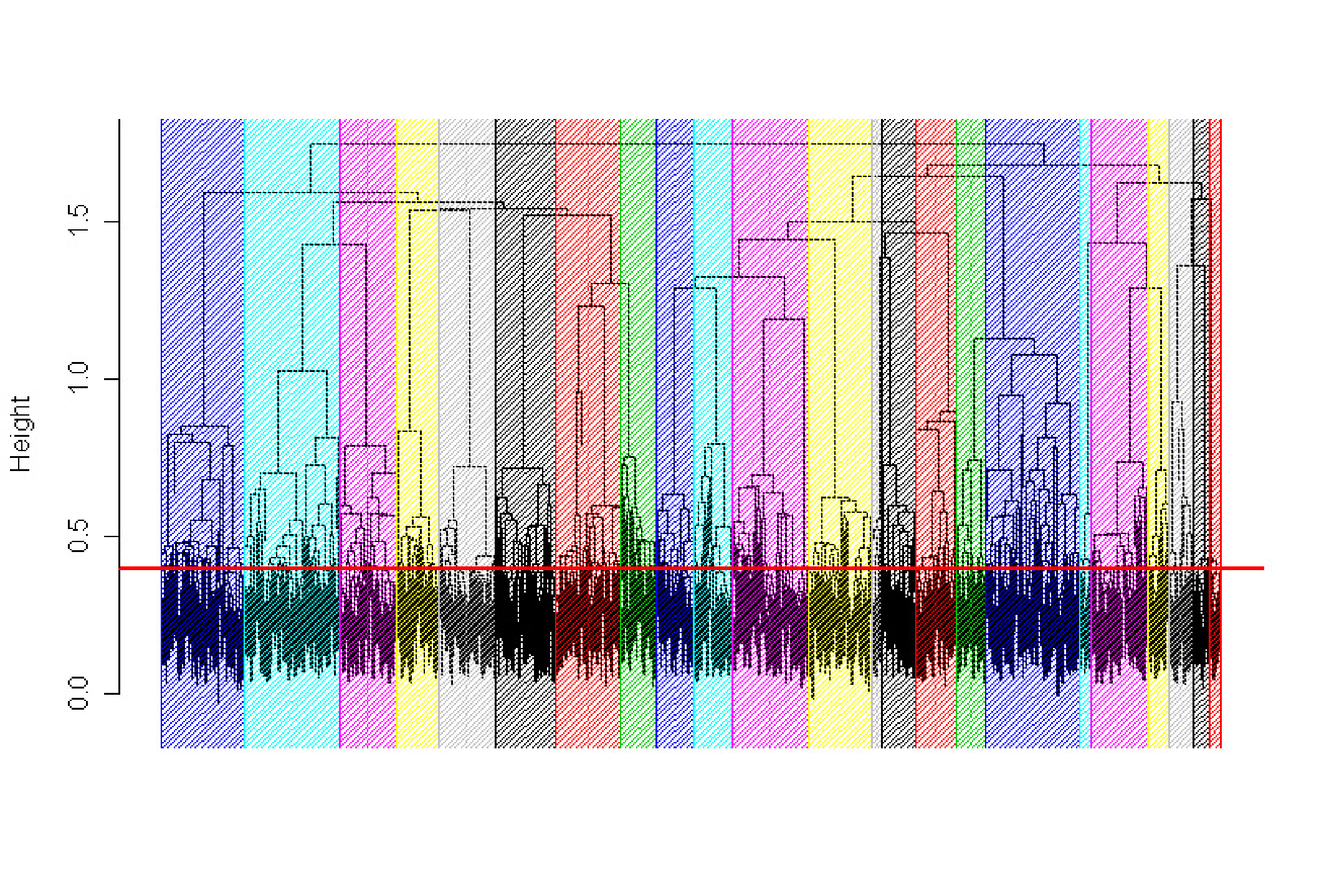}
\caption{Hierarchical tree constructed by FOC. Each leaf represents
one gene/clone on the array. The order of the leaves represents the
order of genes on the genome. The 23 Chromosomes are illustrated
with different colors. Cutting the tree at 0.04 (horizonal red line)
separates the genome into 384 intervals. This cutoff point is chosen
such that no interval contains genes from different
chromosomes.}\label{Fig:app:FOC}
\end{center}
\vspace{-23pt}
\end{figure}

\begin{figure}[h]
\label{figure:heatmap}
\begin{center}
\includegraphics[width=4in,  angle=0]{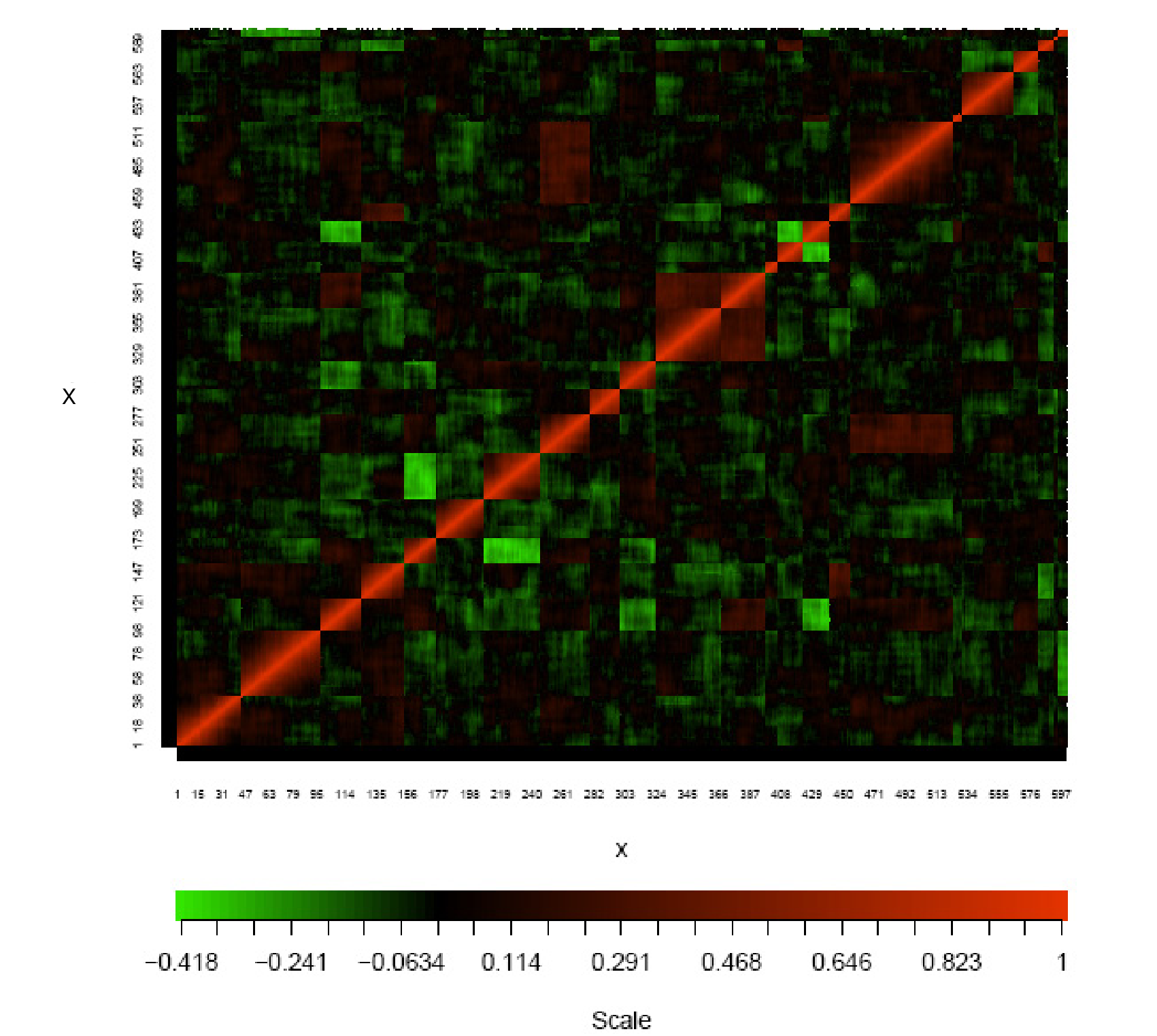}
\includegraphics[width=3.5in,  angle=0]{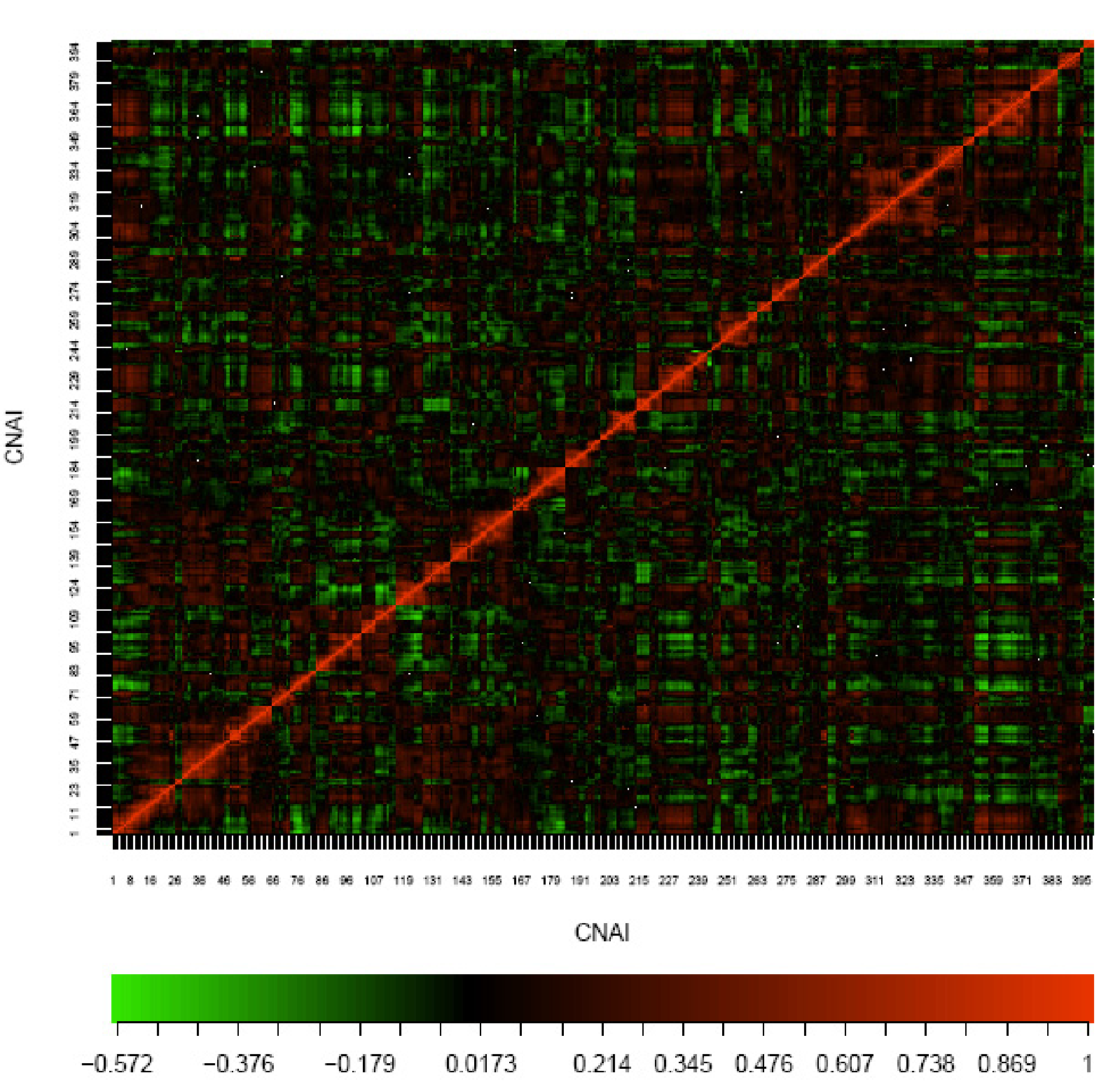}
\end{center}
\caption{Heatmaps of the sample correlations among predictors. Top
panel: simulated data; Bottom panel: real data}
\end{figure}

\begin{figure}[h]
\begin{center}
    \subfigure[\textit{Exp.Net.664}: Inferred network for the 654 breast cancer related genes (based on their expression levels) by
\texttt{space}. Nodes with degrees greater than ten are drawn in
blue.]{
          \label{Fig:spaceNet}
          \includegraphics[width=10cm,
                          angle=0]{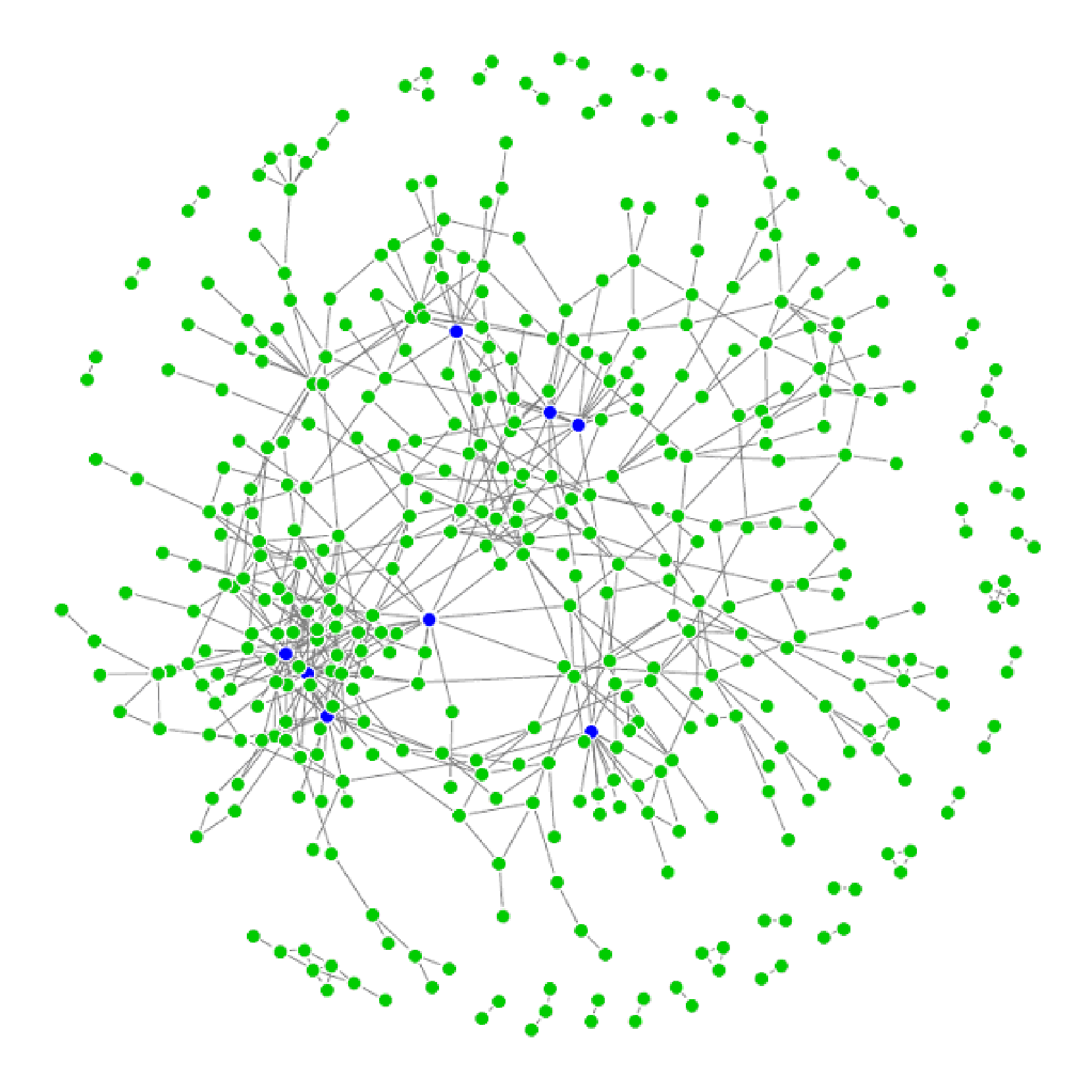}} \hspace{.0in}
   \subfigure[Degree distribution of network  \textit{Exp.Net.664}.]{
          \label{Fig:degree}
          \includegraphics[width=7cm,
          angle=0]{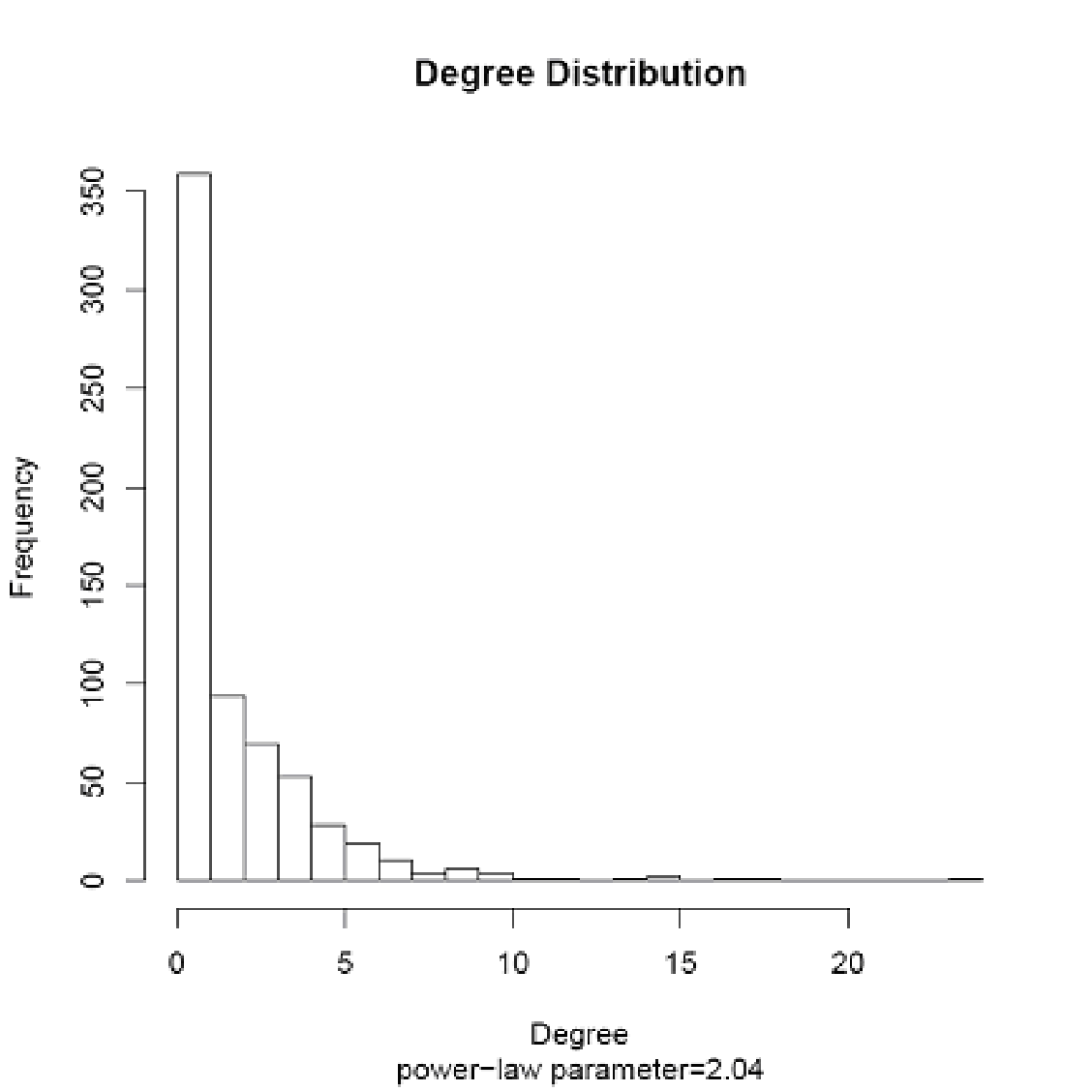}} \hspace{.0in}
\caption{Inferred RNA interaction network.} \label{Fig:Space}
\end{center}
\vspace{-23pt}
\end{figure}

\begin{figure}[h]
\begin{center}
\includegraphics[width=15cm, angle=0]{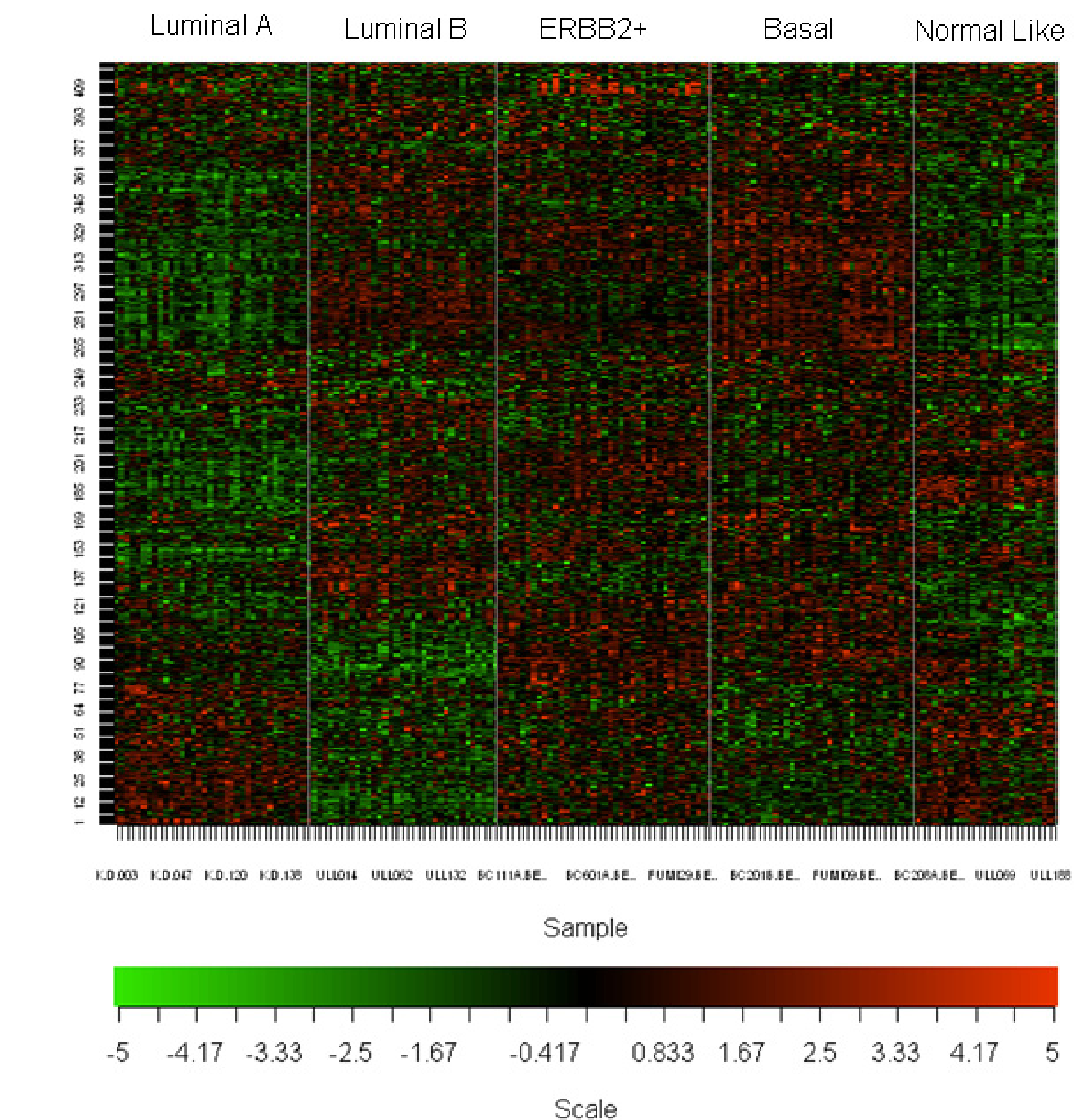}
\caption{Heatmap showing the expressions of the 449 intrinsic genes
in the 172 breast cancer tumor samples. Each column represents one
sample and each row represents one gene. The 172 samples are divided
into 5 clusters (subtypes). From the left to the right, the 5
subtypes are: Luminal Subtype A, Luminal Subtype B,
ERBB2-overexpressing subtype, Basal Subtype and Normal Breast-like
Subtype.}\label{Fig:subtype}
\end{center}
\vspace{-23pt}
\end{figure}

%% file: remMap_042309.bbl
\begin{thebibliography}{}

\bibitem[\protect\citeauthoryear{Albertson et al.}{Albertson et al.}{2003}]{Albertson:2003}
Albertson, D. G., C. Collins, F. McCormick, and J. W. Gray, (2003),
\newblock ``Chromosome aberrations in solid tumors,''
\newblock {\em Nature Genetics\/},~34.

\bibitem[\protect\citeauthoryear{Antoniadis and Fan}{Antoniadis and Fan}{2001}]{Antoniadis:2001} Antoniadis,
A., and Fan, J., (2001), \newblock
``Regularization of wavelet approximations,'' \newblock {\em Journal of the
American Statistical Association}, {96}, 939--967.


\bibitem[\protect\citeauthoryear{Bai and Louh}{Bai and Louh}{2008}]{Bai:2008}
Bai T, Luoh SW., (2008)
\newblock ``GRB-7 facilitates HER-2/Neu-mediated signal transduction and tumor formation,''
\newblock {\em Carcinogenesis\/},~29(3),~473-9.


\bibitem[\protect\citeauthoryear{Bakin}{Bakin}{1999}]{Bakin:1999}
Bakin, S., (1999),
\newblock ``Adaptive regression and model selection in data mining problems,''
\newblock {\em PhD Thesis \/},~Australian National University,~Canberra.

\bibitem[\protect\citeauthoryear{Bedrick et al.}{Bedrick et al.}{1994}]{Bedrick:1994}
Bedrick, E. and Tsai, C.,(1994),
\newblock ``Model selection for multivariate regression in small samples,''
\newblock {\em Biometrics\/},~50,~226每231.

\bibitem[\protect\citeauthoryear{Bergamaschi et al.}{Bergamaschi et al.}{2006}]{Bergamaschi:2006}
Bergamaschi, A., Kim, Y. H., Wang, P., Sorlie, T., Hernandez-Boussard, T., Lonning, P. E., Tibshirani, R., Borresen-Dale, A. L., and Pollack, J. R., (2006),
\newblock ``Distinct patterns of DNA copy number alteration are associated with different clinicopathological features and gene-expression 
subtypes of breast cancer,''
\newblock {\em Genes Chromosomes Cancer\/},~45,~1033-1040.


\bibitem[\protect\citeauthoryear{Bergamaschi et al.}{Bergamaschi et al.}{2008}]{Bergamaschi:2008}
Bergamaschi, A., Kim, Y.H., Kwei, K.A., Choi, Y.L., Bocanegra, M., Langerod, A., Han, W., Noh, D.Y., Huntsman, D.G., Jeffrey, S.S., 
Borresen-Dale, A. L., and Pollack, J.R., (2008),
\newblock ``CAMK1D amplification implicated in epithelial-mesenchymal transition in basal-like breast cancer,''
\newblock {\em Mol Oncol\/},~In Press.



\bibitem[\protect\citeauthoryear{Breiman et al.}{Breiman et al.}{1997}]{Breiman:1997}
Breiman, L. and Friedman, J. H., (1997),
\newblock ``Predicting multivariate responses in multiple linear regression (with discussion),''
\newblock {\em J. R. Statist. Soc. B\/},~59,~3-54.


\bibitem[\protect\citeauthoryear{Brown et al.}{Brown et al.}{1999}]{Brown:1999}
Brown, P., Fearn, T. and Vannucci, M., (1999),
\newblock ``The choice of variables in multivariate regression: a non-conjugate Bayesian decision theory approach,''
\newblock {\em Biometrika\/},~86,~635每648.


\bibitem[\protect\citeauthoryear{Brown et al.}{Brown et al.}{1998}]{Brown:1998}
Brown, P., Vannucci, M. and Fearn, T., (1998),
\newblock ``Multivariate Bayesian variable selection and prediction,''
\newblock {\em J. R. Statist. Soc. B\/},~60,~627每641.


\bibitem[\protect\citeauthoryear{Brown et al.}{Brown et al.}{2002}]{Brown:2002}
Brown, P., Vannucci, M. and Fearn, T.,(2002),
\newblock ``Bayes model averaging with selection of regressors,''
\newblock {\em J. R. Statist. Soc. B\/},~64,~519每536.


\bibitem[\protect\citeauthoryear{Chang et al.}{Chang et al.}{2004}]{Chang:2004}
Chang HY, Sneddon JB, Alizadeh AA, Sood R, West RB, et al., (2004),
\newblock ``Gene expression signature of fibroblast serum response predicts human cancer progression: Similarities between tumors and wounds,''
\newblock {\em PLoS Biol\/},~2(2).


\bibitem[\protect\citeauthoryear{Efron, Hastie, Johnstone, and
 Tibshirani}{Efron et~al.}{2004}]{lars}
Efron, B., Hastie, T., Johnstone, I., and Tibshirani, R. (2004),
\newblock ``Least Angle Regression,''
\newblock {\em Annals of Statistics\/},~32,~407--499.


\bibitem[\protect\citeauthoryear{Frank et al.}{Frank et al.}{1993}]{Frank:1993}
Frank, I. and Friedman, J.,(1993),
\newblock ``A statistical view of some chemometrics regression tools (with discussion),''
\newblock {\em Technometrics\/},~35,~109每148.

\bibitem[\protect\citeauthoryear{Fu}{Fu}{1998}]{Fu:1998}
Fu, W., (1998),
\newblock ``Penalized regressions: the bridge vs the lasso,''
\newblock {\em Journal of Computational and Graphical Statistics\/},~7(3):417-433.

\bibitem[\protect\citeauthoryear{Friedman et al.}{Friedman et al.}{2008}]{Friedman:2008}
Friedman, J., Hastie, T. and Tibshirani, R.,(2008),
\newblock ``Regularized Paths for Generalized Linear Models via Coordinate Descent,''
\newblock {\em Techniqual Report\/},~Department of Statistics,~Stanford University.


\bibitem[\protect\citeauthoryear{Friedman et al.}{Friedman et al.}{2007}]{Friedman:2007}
Friedman, J., Hastie, T. and Tibshirani, R.,(2007),
\newblock ``Pathwise coordinate optimization,''
\newblock {\em The Annals of Applied Statistics.\/},~1(2),~302-332 .

\bibitem[\protect\citeauthoryear{Fujikoshi et al.}{Fujikoshi et al.}{1997}]{Fujikoshi:1997}
Fujikoshi,Y. and Satoh, K., (1997),
\newblock ``Modified AIC and Cp in multivariate linear regression,''
\newblock {\em Biometrika\/},~84,~707每716.


\bibitem[\protect\citeauthoryear{Gardner et al.}{Gardner et al.}{2003}]{Gardner:2003}
Gardner, T. S., D. DI Bernardo, D. Lorenz, and J. J. Collins, (2003)
\newblock ``Inferring genetic networks and identifying compound mode of action via expression profiling,''
\newblock {\em Science\/},~301



\bibitem[\protect\citeauthoryear{Hyman et al.}{Hyman et al.}{2002}]{Hyman:2002}
Hyman, E., P. Kauraniemi, S. Hautaniemi, M. Wolf, S. Mousses, E. Rozenblum, M. Ringner, G. Sauter,
O. Monni, A. Elkahloun, O.-P. Kallioniemi, and A. Kallioniemi, (2002),
\newblock ``Impact of dna amplification on gene expression patterns in breast cancer,''
\newblock {\em Cancer Res\/},~62.


\bibitem[\protect\citeauthoryear{Izenman}{Izenman}{1975}]{Izenman:1975}
Izenman, A., (1975),
\newblock ``Reduced-rank regression for the multivariate linear model,''
\newblock {\em J. Multiv. Anal.\/},~5,~248每264.


\bibitem[\protect\citeauthoryear{Jeong et al.}{Jeong et al.}{2001}]{Jeong:2001}
Jeong, H., S. P. Mason, A. L. Barabasi, and Z. N. Oltvai, (2001),
\newblock ``Lethality and centrality in protein networks,''
\newblock {\em Nature\/},~(411)


\bibitem[\protect\citeauthoryear{Kapp et al.}{Kapp et al.}{2006}]{Kapp:2006}
Kapp, A. V., Jeffrey, S. S., Langerod, A., Borresen-Dale, A. L., Han, W., Noh, D. Y., Bukholm, I. R., Nicolau, M., Brown, P. O. and 
Tibshirani, R., (2006),
\newblock ``Discovery and validation of breast cancer subtypes,''
\newblock {\em BMC Genomics\/},~7,~231.

\bibitem[\protect\citeauthoryear{Kao and Pollack}{Kao and Pollack}{2006}]{Kao:2006}
Kao J, Pollack JR., (2006),
\newblock ``RNA interference-based functional dissection of the 17q12 amplicon in breast cancer reveals contribution of coamplified genes,''
\newblock {\em Genes Chromosomes Cancer\/},~45(8),~761-9.


\bibitem[\protect\citeauthoryear{Kim et al.}{Kim et al.}{2008}]{Kim:2008}
Kim, S., Sohn, K.-A., Xing E. P., (2008)
\newblock ``A multivariate regression approach to association analysis of quantitative trait network''
\newblock {\em http://arxiv.org/abs/0811.2026\/}.


\bibitem[\protect\citeauthoryear{Langerod et al.}{Langerod et al.}{2007}]{Langerod:2007}
Langerod, A., Zhao, H., Borgan, O., Nesland, J. M., Bukholm, I. R., Ikdahl, T., Karesen, R., Borresen-Dale, A. L., Jeffrey, S. S., (2007),
\newblock ``TP53 mutation status and gene expression profiles are powerful
prognostic markers of breast cancer,''
\newblock {\em Breast Cancer Res\/},~9,~R30.

\bibitem[\protect\citeauthoryear{Lutz and B{\"{u}}hlmann}{Lutz and B{\"{u}}hlmann}{2006}]{Lutz:2006}
Lutz, R. and B{\"{u}}hlmann, P., (2006),
\newblock ``Boosting for high-multivariate responses in high-dimensional linear regression,''
\newblock {\em Statist. Sin.\/},~16,~471每494.


\bibitem[\protect\citeauthoryear{Obozinskiy et al.}{Obozinskiy et al.}{2008}]{Obozinskiy:2008}
Obozinskiy, G., Wainwrighty, M.J., Jordany, M. I., (2008)
\newblock ``Union support recovery in high-dimensional multivariate regression,''
\newblock {\em http://arxiv.org/abs/0808.0711\/}.


\bibitem[\protect\citeauthoryear{Paik et al.}{Paik et al.}{2004}]{Paik:2004}
Paik S, Shak S, Tang G, Kim C, Baker J, et al., (2004),
\newblock ``A multigene assay to predict recurrence of tamoxifen-treated, node-negative breast cancer,''
\newblock {\em N Engl J Med\/},~351(27),~2817-2826.

\bibitem[\protect\citeauthoryear{Peng et al.}{Peng et al.}{2008}]{space:2008}
Peng, J., P. Wang, N. Zhou, J. Zhu, (2008),
\newblock ``Partial Correlation Estimation by Joint Sparse Regression Models,''
\newblock {\em JASA\/},~to appear.

\bibitem[\protect\citeauthoryear{Pollack et al.}{Pollack et al.}{2002}]{Pollack:2002}
Pollack, J., T. Srlie, C. Perou, C. Rees, S. Jeffrey, P. Lonning, R. Tibshirani, D. Botstein, A. Brresen-Dale,
and Brown,P., (2002),
\newblock ``Microarray analysis reveals a major direct role of dna copy number alteration in the
transcriptional program of human breast tumors,''
\newblock {\em Proc Natl Acad Sci\/},~99(20).



\bibitem[\protect\citeauthoryear{Reinsel and Velu}{Reinsel and Velu}{1998}]{Reinsel:1998}
Reinsel, G. and Velu, R., (1998),
\newblock ``Multivariate Reduced-rank Regression: Theory and Applications,''
\newblock {\em New York\/},~Springer.




\bibitem[\protect\citeauthoryear{Saal et al.}{Saal et al.}{2007}]{Saal:2007}
Saal LH, Johansson P, Holm K, Gruvberger-Saal SK, She QB, et al., (2007),
\newblock ``Poor prognosis in carcinoma is associated with a gene expression signature of aberrant PTEN tumor suppressor pathway activity,''
\newblock {\em Proc Natl Acad Sci U S A\/},~104(18),~7564-7569.

\bibitem[\protect\citeauthoryear{Sorlie et al.}{Sorlie et al.}{2001}]{Sorlie:2001}
Sorlie, T., Perou, C. M., Tibshirani, R., Aas, T., Geisler, S., Johnsen, H., Hastie, T., Eisen, M. B., van de Rijn, M., Jeffrey, S. S., 
Thorsen, T., Quist, H., Matese,J. C., Brown,P. O., Botstein, D., L鷢ning P. E., and B鷨resen-Dale, A.L., (2001),
\newblock ``Gene expression patterns of breast carcinomas distinguish tumor subclasses with clinical implications,''
\newblock {\em Proc Natl Acad Sci U S A\/},~98,~10869-10874.

\bibitem[\protect\citeauthoryear{Sorlie et al.}{Sorlie et al.}{2003}]{Sorlie:2003}
Sorlie, T., Tibshirani, R., Parker, J., Hastie, T., Marron, J. S., Nobel, A., Deng, S., Johnsen, H., Pesich, R., Geisler, S., 
Demeter,J., Perou,C. M., L鷢ning, P. E., Brown, P. O., B鷨resen-Dale, A.-L., and Botstein, D., (2003),
\newblock ``Repeated observation of breast tumor subtypes in independent gene expression data sets,''
\newblock {\em Proc Natl Acad Sci U S A\/},~100,~8418-8423.

\bibitem[\protect\citeauthoryear{Sotiriou et al.}{Sotiriou et al.}{2006}]{Sotiriou:2006}
Sotiriou C, Wirapati P, Loi S, Harris A, Fox S, et al.,(2006),
\newblock ``Gene expression profiling in breast cancer: Understanding the molecular basis of histologic grade to improve prognosis,''
\newblock {\em J Natl Cancer Inst\/},~98(4),~262-272.




\bibitem[\protect\citeauthoryear{Tibshirani}{Tibshirani}{1996}]{Lasso:1996}
Tibshirani, R., (1996)
\newblock ``Regression shrinkage and selection via the lasso,''
\newblock {\em J. R. Statist. Soc. B \/},~58,~267每288.

\bibitem[\protect\citeauthoryear{Tibshirani and Wang}{Tibshirani and Wang}{2008}]{cghFLasso2008}
Tibshirani, R. and Wang, P., (2008)
\newblock ``Spatial smoothing and hot spot detection for cgh data using the fused lasso,''
\newblock {\em Biostatistics \/},~9(1),~18-29.

\bibitem[\protect\citeauthoryear{Turlach et al.}{Turlach et al.}{2005}]{Turlach:2005}
Turlach, B., Venables, W. and Wright, S.,(2005),
\newblock ``Simultaneous variable selection,''
\newblock {\em Technometrics\/},~47,~349每363.


\bibitem[\protect\citeauthoryear{Wang}{Wang}{2004}]{Wang:thesis}
Wang, P., (2004)
\newblock ``Statistical methods for CGH array analysis,'' 
\newblock {\em Ph.D. Thesis, Stanford University\/}.

\bibitem[\protect\citeauthoryear{Wang et al.}{Wang et al.}{2005}]{Wang:2005}
Wang Y, Klijn JG, Zhang Y, Sieuwerts AM, Look MP, et al.,(2005),
\newblock ``Gene-expression profiles to predict distant metastasis of lymph-node-negative primary breast cancer,''
\newblock {\em Lancet\/},~365(9460),~671-679.




\bibitem[\protect\citeauthoryear{van de Vijver et al.}{van de Vijver et al.}{2002}]{vandeVijver:2002}
van de Vijver MJ, He YD, van't Veer LJ, Dai H, Hart AA, Voskuil DW, Schreiber GJ, Peterse JL, Roberts C, Marton MJ, Parrish M, Atsma D, Witteveen A, Glas A, 
Delahaye L, van der Velde T, Bartelink H, Rodenhuis S, Rutgers ET, Friend SH, Bernards R, (2002)
\newblock ``A gene-expression signature as a predictor of survival in breast cancer,''
\newblock {\em N Engl J Med\/},~347(25),~1999-2009.


\bibitem[\protect\citeauthoryear{Yuan et al.}{Yuan et al.}{2007}]{YuanMulti:2007}
Yuan, M., Ekici, A., Lu, Z., and Monterio, R., (2007) 
\newblock ``Dimension reduction and coefficient estimation in multivariate linear regression,''
\newblock {\em J. R. Statist. Soc. B\/},~69(3),~329每346.

\bibitem[\protect\citeauthoryear{Yuan and Lin}{Yuan and Lin}{2006}]{YuanLin:2006}
Yuan, M. and Lin, Y., (2006)
\newblock ``Model Selection and Estimation in Regression with Grouped Variables,''
\newblock {\em Journal of the Royal Statistical Society, Series B,\/},~68(1),~49-67.


\bibitem[\protect\citeauthoryear{Zhao et al.}{Zhao et al.}{2004}]{Zhao:2004}
Zhao, H., Langerod, A., Ji, Y., Nowels, K. W., Nesland, J. M., Tibshirani, R., Bukholm, I. K., Karesen, R., Botstein, D., Borresen-Dale, A. L., 
and Jeffrey, S. S., (2004),
\newblock ``Different gene expression patterns in invasive lobular and ductal carcinomas of the breast,''
\newblock {\em Mol Biol Cell\/},~15,~2523-2536.


\bibitem[\protect\citeauthoryear{Zhao et al.}{Zhao et al.}{2006}]{Zhao:2006}
Zhao, P., Rocha, G., and Yu, B., (2006), \newblock ``Grouped and hierarchical model selection
through composite absolute penalties,'' \newblock {\em Annals of
  Statistics}.  Accepted.

\bibitem[\protect\citeauthoryear{Zou et al.}{Zou et al.}{2007}]{Zou:2007}
Zou, H., Trevor, H. and Tibshirani, R., (2007),
\newblock ``On degrees of freedom of the lasso,''
\newblock {\em Annals of Statistics\/},~35(5),~2173-2192.


\bibitem[\protect\citeauthoryear{Zou et al.}{Zou et al.}{2005}]{Zou:2005}
Zou, H. and Trevor, T., (2005),
\newblock ``Regularization and Variable Selection via the Elastic Net,''
\newblock {\em Journal of the Royal Statistical Society, Series B\/},~67(2),~301-320.




\end{thebibliography}

\begin{thebibliography}{}


\bibitem[\protect\citeauthoryear{Chang et al.}{Chang et al.}{2004}]{Chang:2004}
Chang HY, Sneddon JB, Alizadeh AA, Sood R, West RB, et al., (2004),
\newblock ``Gene expression signature of fibroblast serum response predicts human cancer progression: Similarities between tumors and wounds,''
\newblock {\em PLoS Biol\/},~2(2).




\bibitem[\protect\citeauthoryear{Dowdy et al.}{Dowdy et al.}{2005}]{Dowdy:2005}
Dowdy SC, Gostout BS, Shridhar V, Wu X, Smith DI, Podratz KC, Jiang SW., (2005)
\newblock ``Biallelic methylation and silencing of paternally expressed gene 3(PEG3) in gynecologic cancer cell lines,''
\newblock {\em Gynecol Oncol\/},~99(1),~126-34.


\bibitem[\protect\citeauthoryear{Efron, Hastie, Johnstone, and
 Tibshirani}{Efron et~al.}{2004}]{lars}
Efron, B., Hastie, T., Johnstone, I., and Tibshirani, R. (2004),
\newblock ``Least Angle Regression,''
\newblock {\em Annals of Statistics\/},~32,~407--499.




\bibitem[\protect\citeauthoryear{Johnson et al.}{Johnson et al.}{2002}]{Johnson:2002}
Johnson MD, Wu X, Aithmitti N, Morrison RS., (2002)
\newblock ``Peg3/Pw1 is a mediator between p53 and Bax in DNA damage-induced neuronal death,''
\newblock {\em J Biol Chem\/},~277(25),~23000-7.




\bibitem[\protect\citeauthoryear{Koh et al.}{Koh et al.}{2005}]{Koh:2005}
Koh WP, Yuan JM, Van Den Berg D, Lee HP, Yu MC., (2005)
\newblock ``Polymorphisms in angiotensin II type 1 receptor and angiotensin I-converting enzyme genes and breast 
cancer risk among Chinese women in Singapore,''
\newblock {\em Carcinogenesis\/},~26(2),~459-64.


\bibitem[\protect\citeauthoryear{Kremmidiotis et al.}{Kremmidiotis et al.}{1998}]{Kremmidiotis:1998}
Kremmidiotis G, Baker E, Crawford J, Eyre HJ, Nahmias J, Callen DF., (1998),
\newblock ``Localization of human cadherin genes to chromosome regions exhibiting cancer-related loss of heterozygosity,''
\newblock {\em Genomics\/},~49(3),~467-71.


\bibitem[\protect\citeauthoryear{Newman}{Newman}{2003}]{Newman:2003}
Newman M, (2003),
\newblock ``The Structure and Function of Complex Networks,''
\newblock {\em Society
for Industrial and Applied Mathematics\/},~45(2),~167-256.


\bibitem[\protect\citeauthoryear{Paik et al.}{Paik et al.}{2004}]{Paik:2004}
Paik S, Shak S, Tang G, Kim C, Baker J, et al., (2004),
\newblock ``A multigene assay to predict recurrence of tamoxifen-treated, node-negative breast cancer,''
\newblock {\em N Engl J Med\/},~351(27),~2817-2826.


\bibitem[\protect\citeauthoryear{Peng et al.}{Peng et al.}{2008}]{space:2008}
Peng, J., P. Wang, N. Zhou, J. Zhu, (2008),
\newblock ``Partial Correlation Estimation by Joint Sparse Regression Models,''
\newblock {\em JASA\/},~to appear.




\bibitem[\protect\citeauthoryear{Rizki et al.}{Rizki et al.}{2007}]{Rizki:2007}
Rizki A, Mott JD, Bissell MJ, (2007)
\newblock ``Polo-like kinase 1 is involved in invasion through extracellular matrix,''
\newblock {\em Cancer Res\/},~67(23),~11106-10.



\bibitem[\protect\citeauthoryear{Saal et al.}{Saal et al.}{2007}]{Saal:2007}
Saal LH, Johansson P, Holm K, Gruvberger-Saal SK, She QB, et al., (2007),
\newblock ``Poor prognosis in carcinoma is associated with a gene expression signature of aberrant PTEN tumor suppressor pathway activity,''
\newblock {\em Proc Natl Acad Sci U S A\/},~104(18),~7564-7569.

\bibitem[\protect\citeauthoryear{Sorlie et al.}{Sorlie et al.}{2003}]{Sorlie:2003}
Sorlie, T., Tibshirani, R., Parker, J., Hastie, T., Marron, J. S., Nobel, A., Deng, S., Johnsen, H., Pesich, R., Geisler, S., 
Demeter,J., Perou,C. M., Lønning, P. E., Brown, P. O., Børresen-Dale, A.-L., and Botstein, D., (2003),
\newblock ``Repeated observation of breast tumor subtypes in independent gene expression data sets,''
\newblock {\em Proc Natl Acad Sci U S A\/},~100,~8418-8423.


\bibitem[\protect\citeauthoryear{Sotiriou et al.}{Sotiriou et al.}{2006}]{Sotiriou:2006}
Sotiriou C, Wirapati P, Loi S, Harris A, Fox S, et al.,(2006),
\newblock ``Gene expression profiling in breast cancer: Understanding the molecular basis of histologic grade to improve prognosis,''
\newblock {\em J Natl Cancer Inst\/},~98(4),~262-272.



\bibitem[\protect\citeauthoryear{Talvinen et al.}{Talvinen et al.}{2008}]{Talvinen:2008}
Talvinen K, Tuikkala J, Nevalainen O, Rantanen A, Hirsim\"{a}ki P, Sundstr\"{o}m J, Kronqvist P, (2008),
\newblock ``Proliferation marker securin identifies favourable outcome in invasive ductal breast cancer,''
\newblock {\em Br J Cancer\/},~99(2),~335-40.

\bibitem[\protect\citeauthoryear{Thomassen et al.}{Thomassen et al.}{2008}]{Thomassen:2008}
Thomassen M, Tan Q, Kruse TA, (2008)
\newblock ``Gene expression meta-analysis identifies chromosomal regions and candidate genes involved in breast cancer metastasis,''
\newblock {\em Breast Cancer Res Treat\/},~Feb 22,~Epub.

\bibitem[\protect\citeauthoryear{Tibshirani and Wang}{Tibshirani and Wang}{2008}]{cghFLasso2008}
Tibshirani, R. and Wang, P., (2008)
\newblock ``Spatial smoothing and hot spot detection for cgh data using the fused lasso,''
\newblock {\em Biostatistics \/},~9(1),~18-29.


\bibitem[\protect\citeauthoryear{van de Vijver et al.}{van de Vijver et al.}{2002}]{vandeVijver:2002}
van de Vijver MJ, He YD, van't Veer LJ, Dai H, Hart AA, Voskuil DW, Schreiber GJ, Peterse JL, Roberts C, Marton MJ, Parrish M, Atsma D, Witteveen A, Glas A, 
Delahaye L, van der Velde T, Bartelink H, Rodenhuis S, Rutgers ET, Friend SH, Bernards R, (2002)
\newblock ``A gene-expression signature as a predictor of survival in breast cancer,''
\newblock {\em N Engl J Med\/},~347(25),~1999-2009.


\bibitem[\protect\citeauthoryear{Voduc et al.}{Voduc et al.}{2008}]{Voduc:2008}
Voduc D, Cheang M, Nielsen T, (2008),
\newblock ``GATA-3 expression in breast cancer has a strong association with estrogen receptor but lacks independent prognostic value,''
\newblock {\em Cancer Epidemiol Biomarkers Prev\/},~17(2),~365-73.



\bibitem[\protect\citeauthoryear{Wang}{Wang}{2004}]{Wang:thesis}
Wang, P., (2004)
\newblock ``Statistical methods for CGH array analysis,'' 
\newblock {\em Ph.D. Thesis, Stanford University\/},~80-81.

\bibitem[\protect\citeauthoryear{Wang et al.}{Wang et al.}{2005}]{CLAC:2005}
Wang P, Kim Y, Pollack J, Narasimhan B, Tibshirani R, (2005)
\newblock ``A method for calling gains and losses in array CGH data,''
\newblock {\em Biostatistics}, ~6(1),~45-58. 

\bibitem[\protect\citeauthoryear{Wang et al.}{Wang et al.}{2005}]{Wang:2005}
Wang Y, Klijn JG, Zhang Y, Sieuwerts AM, Look MP, et al.,(2005),
\newblock ``Gene-expression profiles to predict distant metastasis of lymph-node-negative primary breast cancer,''
\newblock {\em Lancet\/},~365(9460),~671-679.


\bibitem[\protect\citeauthoryear{Yuan and Lin}{Yuan and Lin}{2006}]{YuanLin:2006}
Yuan, M. and Lin, Y., (2006)
\newblock ``Model Selection and Estimation in Regression with Grouped Variables,''
\newblock {\em Journal of the Royal Statistical Society, Series B,\/},~68(1),~49-67.


\end{thebibliography}
